\documentclass[12pt,preprint]{aastex}
\usepackage{xspace}
\usepackage{graphicx}
\usepackage{natbib}
\usepackage{arydshln}
\newcommand{\nraoblurb}{The National Radio Astronomy Observatory is
a facility of the National Science Foundation operated under cooperative
agreement by Associated Universities, Inc.}
\newcommand{\hide}[1]{}

\newcommand{\gsim}{\ensuremath{\,\gtrsim\,}\xspace}
\newcommand{\lsim}{\ensuremath{\,\lesssim\,}\xspace}
\newcommand{\gl}{\ensuremath{\ell}\xspace}
\newcommand{\gb}{\ensuremath{{\it b}}\xspace}
\newcommand{\absb}{\ensuremath{\vert\,\gb\,\vert}\xspace}

\newcommand{\lb}{\ensuremath{(\gl,\gb)}\xspace}
\newcommand{\lv}{\ensuremath{(\gl,v)}\xspace}

\newcommand{\lbv}{\ensuremath{(\gl,\gb,v)}\xspace}

\newcommand{\kms}{\ensuremath{\,{\rm km\,s^{-1}}}\xspace}

\newcommand{\cm}{\ensuremath{\,{\rm cm}}\xspace}

\newcommand{\kpc}{\ensuremath{\,{\rm kpc}}\xspace}
\newcommand{\K}{\ensuremath{\,{\rm K}}\xspace}

\newcommand{\ghz}{\ensuremath{\,{\rm GHz}}\xspace}

\newcommand{\degree}{\ensuremath{\,^\circ}\xspace}

\newcommand{\jy}{\ensuremath{\,{\rm Jy}}\xspace}

\newcommand{\mjy}{\ensuremath{\,{\rm mJy}}\xspace}

\newcommand{\te}{\ensuremath{{T_{e}}}\xspace}

\newcommand{\rgal}{\ensuremath{\,R_G}\xspace}   

\newcommand{\hi}{{\rm H\,{\footnotesize I}}\xspace}
\newcommand{\hii}{{\rm H\,{\footnotesize II}}\xspace}

\newcommand{\hal}[1]{\ensuremath{{\rm H}\,{#1}\,\alpha\xspace}}
\newcommand{\hna}{\ensuremath{{\rm H}\,{\rm n}\,\alpha\xspace}}
\newcommand{\hnaa}{\ensuremath{\langle\,\hna\,\rangle\xspace}}

\newcommand{\water}{\ensuremath{\rm H_2O}\xspace}

\shorttitle{Finding Distant Galactic HII Regions}
\shortauthors{Anderson et al.}

\begin{document}

\title{Finding Distant Galactic HII Regions}

\author{L.~D.~Anderson\altaffilmark{1,2},
  W.~P.~Armentrout\altaffilmark{1}, B.~M.~Johnstone\altaffilmark{1},
  T.~M.~Bania\altaffilmark{3}, Dana~S.~Balser\altaffilmark{4},
  Trey~V.~Wenger\altaffilmark{5}, V.~Cunningham\altaffilmark{1}}

\altaffiltext{1}{Department of Physics and Astronomy, West Virginia University, Morgantown WV 26506, USA}
\altaffiltext{2}{Adjunct Astronomer at the National Radio
  Astronomy Observatory, P.O. Box 2, Green Bank, WV 24944, USA}
\altaffiltext{3}{Institute for Astrophysical Research, Department of Astronomy, Boston University, 725 Commonwealth Ave., Boston MA 02215, USA}
\altaffiltext{4}{National Radio Astronomy Observatory, 520 Edgemont Road, Charlottesville VA, 22903-2475, USA}
\altaffiltext{5}{Astronomy Department, University of Virginia, P.O. Box 3818, Charlottesville VA 22903-0818, USA}




%
%
%






\begin{abstract}
The WISE Catalog of Galactic \hii\ Regions contains $\sim2000$
\hii\ region candidates lacking ionized gas spectroscopic observations.
All candidates have the characteristic \hii\ region mid-infrared
morphology of WISE $12\,\micron$ emission surrounding $22\,\micron$
emission, and additionally have detected radio continuum emission.  We here
report Green Bank Telescope (GBT) hydrogen radio recombination line (RRL) and radio continuum
detections at X-band (9\,GHz; 3\,cm) of 302 WISE \hii\ region
candidates (out of 324 targets observed) in the zone $225\degree
\ge \ell \ge -20\degree$, $\absb \le 6\degree$.  Here we extend the
sky coverage of our \hii\ region Discovery Survey (HRDS), which now
contains nearly 800 \hii\ regions distributed across the entire northern sky.  We
provide LSR velocities for the 302 detections and kinematic
distances for 131 of these.  Of the 302 new detections,
five have \lbv coordinates consistent with the Outer Scutum-Centaurus
Arm (OSC), the most distant molecular spiral arm of the Milky Way.
Due to the Galactic warp, these nebulae are found at Galactic
latitudes $>1\degree$ in the first Galactic quadrant, and therefore
were missed in previous surveys of the Galactic plane.  One additional
region has a longitude and velocity consistent with the OSC but lies
at a negative Galactic latitude (G039.183$-$01.422; $-54.9\,\kms$).
With Heliocentric distances $>22$\,kpc and Galactocentric distances
$>16$\,kpc, the OSC \hii\ regions are the most distant known in the
Galaxy.  We detect an additional three \hii\ regions near $\ell \simeq
150\degree$ whose LSR velocities place them at Galactocentric radii
$>19$\,kpc.  If their distances are correct, these nebulae may
represent the limit to Galactic massive star formation.
\end{abstract}


\section{Introduction\label{sec:intro}}

Galactic \hii\ regions are the formation sites of massive OB stars.
Due to their short lifetimes, \hii\ regions locate star formation in
the present epoch and are therefore good tracers of Galactic spiral
structure.  \hii\ regions are extremely bright at infrared and radio
wavelengths and can be seen across the entire Galactic disk. It is
perhaps surprising that the census of Galactic \hii\ regions remains
vastly incomplete.  We have found that by combining \hii\ region
candidates identified at infrared (IR) wavelengths with radio
spectroscopic observations, we can uncover large populations of
heretofore unknown \hii\ regions and determine their locations in the
Galaxy.

The Green Bank Telescope \hii\ Region Discovery Survey \citep[GBT
  HRDS;][]{bania10} detected radio recombination line (RRL) emission
from 448 previously unknown \hii\ regions at X-band (9\ghz; 3\cm)
\citep[][hereafter A11]{anderson11}.  The GBT HRDS extends over 168
square degrees, from $67\,\arcdeg \ge\ \gl \ge\ -17\,\arcdeg$ with
$\absb \le 1\,\arcdeg$.  Since the ISM is optically thin at
cm-wavelengths, RRL emission can be detected from \hii\ regions across the
entire Galactic disk. The HRDS targets were selected based on
spatially coincident 24\,\micron\ \citep[Spitzer MIPSGAL;][]{carey09}
and 21\,cm continuum \citep[VGPS;][]{stil06} emission.  The HRDS
doubled the number of known \hii\ regions in its survey zone.  We
followed this effort using the Arecibo Observatory and detected the
RRL emission from an additional 37 sources \citep{bania12} in the zone
$66\degree \ge \ell \ge 31\degree$; $\absb \le 1\degree$.

Because the HRDS and the Arecibo HRDS extension used the {\it Spitzer}
MIPSGAL survey to identify targets, it was limited to within
$1\arcdeg$ of the Galactic plane, and to within $\sim65\arcdeg$ of the
Galactic center.  The HRDS had the sensitivity to discover nebulae
ionized by O-stars located beyond the Solar orbit on the far side of
the Milky Way (A11).  Due to the Galactic warp, however, the most distant
candidates are found off the Galactic plane, in regions of the Galaxy
not imaged by MIPSGAL.  Further, targets at the largest Galactocentric
radii are generally found in the second and third Galactic quadrants.  These
candidates were not observed in the original HRDS.

The sensitive, high resolution mid-infrared (MIR) data required to
identify new \hii\ regions outside of the MIPSGAL range did not exist
until recently.  This changed with the release of data from the
Wide-field Infrared Survey Explorer (WISE).  WISE has photometric
bands centered at 12\,\micron\ and 22\,\micron, providing spectral
coverage analogous to the 8.0\,\micron\ and 24\,\micron\ {\it Spitzer}
bands.  Using these WISE data, \citet[][hereafter A14]{anderson14}
created a catalog of over 8000 \hii\ regions and candidates spanning
all Galactic longitudes within $8\degree$ of the Galactic midplane.
All these objects have the MIR emission identified in A11 as being
characteristic of \hii\ regions.  About 2000 cataloged objects also
have spatially coincident radio continuum emission, and as A11 found,
this makes them strong candidates for being {\it bona fide}
\hii\ regions.  Because of its sensitivity and all-sky coverage, WISE can
detect the MIR emission from all Galactic \hii\ regions ionized by
single O-stars (A14).

By extending the latitude coverage of the \hii\ region surveys, we can
detect extremely distant massive star formation regions, in particular
in the Outer Scutum-Centaurus Arm (OSC).  The OSC is the most distant
molecular spiral arm known, but it is little studied because it is
located primarily at Galactic latitudes $>1\degree$ in the first
Galactic quadrant \citep{dame11}.  Aside from S83 (G55.11+2.4), no
massive star formation was known in the OSC before the HRDS.  The OSC
arm appears to be the Outer Galaxy continuation of the
Scutum-Centaurus Arm, and is $\sim15$\,kpc from the Galactic center at
a Heliocentric distance of $\sim 20\,\kpc$.

Here we provide a catalog of RRL and radio continuum properties for
302 \hii\ regions detected out of 324 targets identified in the WISE
Catalog of Galactic \hii\ regions.

\section{Observations and Data Analysis\label{sec:data}}
We draw our targets from the MIR objects in the WISE catalog of A14.
Most targets are identified in the WISE catalog as ``candidates'' that
have detected radio continuum emission, but we also include in our
sample Sharpless \hii\ regions \citep{sharpless59} that have never
been observed in RRL emission.  All Sharpless regions observed here
were observed in H$\alpha$ by \citet{fich90}, but at a spectral
resolution of just 15\,\kms.  This low spectral resolution may lead to
inaccurate line parameters (see Section~\ref{sec:sharpless}).  We used
the entire zone of the sky visible by the GBT, north of $-42\arcdeg$
declination.  This corresponds to approximately $270\,\arcdeg > \ell >
-20\,\arcdeg$ at $b = 0\arcdeg$, although we did not detect any
\hii\ regions within $270\degree > \ell > 225\degree$.  All targets
have peak radio continuum emission of at least 30\,mJy (extrapolated
to X-band [3\,cm] wavelengths assuming optically thin free-free continuum emission,
$S_\nu\propto \nu^{-0.1}$).  This flux limit is considerably less than
that of the original HRDS, 70\,mJy.  This results in a number of
target candidates within the original HRDS longitude and latitude
zone.  To get the extrapolated X-band flux densities, we measure the
peak continuum flux densities using the VLA Galactic Plane Survey
continuum data at 21\,cm \citep[VGPS;][]{stil06}, the Canadian
Galactic Plane Survey continuum data at 21\,cm
\citep[CGPS;][]{taylor03}, the NRAO VLA Sky Survey at 20\,cm
\citep[NVSS;][]{condon98}, the SuperMongolo Sky Survey at 36\,cm
\citep[SUMSS;][]{bock99}, and the Southern Galactic Plane Survey
\citep[SGPS;][]{mcclure05} at 21\,cm, depending on the candidate \lb\ position.

For $65\arcdeg \ge \ell \ge 25\degree$, $b \ge
1\arcdeg$, we include in our target list all \hii\ region candidates
with detected radio continuum emission, regardless of their flux
densities.  This is the \lb\ zone of the Outer
Scutum-Centaurus arm \citep{dame11} (see above).  Sources in this
Galactic region have a higher likelihood of being extremely distant and
therefore fainter than the rest of the sample.

We followed the same GBT observational procedure as in the original
HRDS: interleaving total power spectral line (using the
AutoCorrelation Spectrometer [ACS]) and total power continuum (using
the Digital Continuum Receiver [DCR]) observations for each candidate
\hii\ region.  Our observations were made with the GBT 100\,m
telescope from July~2012 through August~2014.  There are seven RRLs
that can be cleanly observed simultaneously with the GBT at X-band:
\hal{87} to \hal{93}.  We average these seven RRLs (each at two
  orthogonal polarizations) to create a single average RRL spectrum.

For the RRL observations, we call each set of total power observations
a ``pair.''  These pairs consist of 6\,min on- and 6\,min off-source
integrations, and the off-source position tracks the same path on the
sky as the on-source position.  In the original HRDS, a single
position-switched pair had an average r.m.s. noise of $\sim1\,\mjy$
after smoothing to 1.86\,\kms\ (see below), and the r.m.s. noise was
$\sim0.7\,\mjy$ for two pairs.  Assuming a line-to-continuum intensity
ratio of 0.1 \citep[][]{quireza06a}, our 30\,mJy cutoff corresponds to
a $3\sigma$ detection in a single pair.  We observe a single pair for
all sources and do additional pairs as needed.

As in the HRDS, the continuum observations consisted of four cross
scans centered on the nominal source position: forward and backward in
Right Ascension (RA) and forward and backward in Declination (Decl.).
Each scan was $40\,\arcmin$ in length and had a slew rate of
$80\,\arcsec$ per second.  We observed at a center frequency of
8665\,MHz with a 320\,MHz bandwidth.  We focused the telescope and
established local pointing corrections on average every two hours
using standard X-band pointing sources.  

In A11, observations of flux calibrators shows that the intensity
scale was accurate at the 10\% level for RRL and continuum data.  
  We established the calibration of the intensity scale using noise
  diodes fired during data acquisition.  We repeat the same
calibration procedure for these data, using 80\,MHz continuum
cross-scans in the RA and Decl. directions of the calibrator 3C147
using the DCR at frequencies from 8.04-9.81\,\ghz.  At 8.3\,GHz
the peak intensity of 3C147 averaged over both polarizations and all
cross scans is 9.81\,K, or 4.90\,Jy using the GBT X-band gain of
2\,K/Jy. This value is within 5\% of the peak flux density at 8.3\,GHz
for 3C147 given in \citet{peng00}.  The flux densities at the
  other continuum frequencies agree to within 10\% of that expected
  for 3C147 assuming a spectral index, $\alpha$, of $0.91\pm0.03$,
  where $S_\nu \propto \nu^{-\alpha}$ \citep{peng00}.  The X-band
intensity scale is relatively insensitive to opacity and elevation
gain corrections, which both have magnitudes of $\lsim 5\%$
\citep{ghigo01}.  Because we are making a discovery survey, we
did not correct our data based on the intensity of 3C147, nor did we
correct for gain and opacity variations. We therefore estimate, as in
the original HRDS, that the intensities and flux densities given in
this paper are uncertain at the 10\% level.



We follow the same data reduction steps as in the original HRDS.  We
again use the TMBIDL software package,
V7.1\footnote{https://github.com/tvwenger/tmbidl.git}.  For the RRL
data, we average the seven RRL transitions with two polarizations to
create a single \hnaa\ spectrum \citep[see][]{balser06}.  Averaging
the spectra in this way improves the RRL signal-to-noise ratio,
allowing for significantly reduced integration times.  We smooth each
combined \hnaa\ spectrum with a normalized Gaussian over five channels
to give a \hnaa\ spectrum with a velocity resolution of 1.86\,\kms.
The original HRDS found hydrogen line widths of $\sim25\,\kms$ on
average, giving us $\gsim10$ independent spectral channels per line
after this smoothing.  We remove a polynomial baseline, typically of
third-order, and fit Gaussians to each detected hydrogen line
component individually for each source.  As in the HRDS, we assume
that the brightest line is from hydrogen.\footnote{There are, however,
  a few sources that have carbon RRLs brighter than those of hydrogen.
  These are easily distinguished because the carbon lines are $\lsim
  10\,\kms$ wide \citep{wenger13}, and are offset $\sim-150\,\kms$
  from the hydrogen lines.}
Thus, we
derive the LSR\footnote{We used the kinematic local standard of rest
  (LSR) frame with the radio definition of the Doppler shift.  The
  kinematic LSR is defined by a solar motion of 20.0\,\kms\ toward
  $(\alpha, \delta) = (18^{\rm h}, +30\degree) [1900.0]$
  \citep{gordon76}.}  velocity, line intensity, and FWHM line width for
each hydrogen RRL component.  We did not correct for the changing beam
size with frequency and therefore the line parameters are averages
over beam sizes from $73\,\arcsec$ to $89\,\arcsec$.

We average the forward and backward continuum scans for each source
(after first flipping the backwards scans) to create a RA
scan and a Decl. scan.  We remove a polynomial baseline, usually
a second-order, from the two scans and then fit Gaussians, again on a
source-by-source basis.  Some sources have multiple emission
components, which we fit with multiple Gaussians.  Using WISE and
radio continuum data as a guide, we attempt to associate a single
Gaussian component with the source observed (see A11 for a more
complete discussion).  As A11 note, source confusion is a significant
issue and the continuum parameters should be used with caution.

We show in Figure~\ref{fig:survey_examples} for three example
  \hii\ regions the WISE three color images (22\,\micron\ in red,
  12\,\micron\ in green, and 3.6\,\micron\ in blue), \hnaa\ spectra,
  and continuum cross scans.  These three regions are representative
of the sample as a whole in terms of angular size, spectral line
intensity, and continuum data quality.

\section{The Catalog of WISE-Identified H\,{\bf\footnotesize II} Regions\label{sec:catalog}}
We detect RRL emission from 302 of the 324 targets,
over 93\% of our sample.  Excluding the 31 regions in the \lb\ zone of
the OSC ($65\degree > \ell > 25\degree; b \ge 1\degree$) where our
flux density threshold was relaxed, these numbers change to 280 of 293
(96\%).  In both cases, these values are similar to the 95\% detection
rate of A11.
The hydrogen RRL data are given in Table~\ref{tab:line}, which lists
the source name, the Galactic longitude and latitude, the line
intensity, the FWHM line width, the LSR velocity, and the r.m.s.
noise in the spectrum.  The errors given in Table \ref{tab:line} for
the line parameters are the $1\,\sigma$ uncertainties from the
Gaussian fits.  For sources with multiple velocity components detected
along the line of sight, we append to their names additional letters
``a'', ``b'', or ``c'' in order of decreasing peak line intensity.
For multiple-velocity sources, if we were able to determine which
  component stems from the discrete \hii\ region (see
  Section~\ref{sec:multvel}) it is flagged with an asterisk
  in the final column of Table~\ref{tab:line}.


We give the radio continuum data in Table \ref{tab:continuum},
which lists the source name, the Galactic longitude and latitude, the
peak intensity in the RA and Decl. scan directions, the FWHM angular
size in the RA and Decl. directions, the integrated flux density, and
the notes on continuum morphology (see below).  We compute the integrated flux density
as in A11:
\begin{equation}
S_i = S_p \, \left(\frac{\theta_{\rm RA}}{\theta_b}\right)
\left(\frac{\theta_{\rm Decl}}{\theta_b}\right)\,,
\label{eq:integrated_flux}
\end{equation}
where $S_p$ is the peak flux found from the peak antenna
  temperature assuming a gain of 2$\K\jy^{-1}$, $\theta_b$ is the GBT beam
size at 8665\,MHz ($87\arcsec$), and $\theta_{\rm RA}$, $\theta_{\rm Decl}$ are the FWHM
angular sizes derived from the RA and Decl. scans
\citep[see][]{kuchar97}.  As in A11, for sources better approximated
by multiple Gaussians, the angular size listed is the maximum extent
of the multi-component composite source and the peak intensity
(e.g., $T_{\rm \alpha}$) is the area under all components
divided by this angular size.  Such ``complex'' objects are marked
with a ``C'' label in the Notes column of Table \ref{tab:continuum}.
The errors given for the continuum parameters are $1\,\sigma$
uncertainties in the Gaussian fits.  We flag nebulae whose peak
continuum emission is within 10\,$\arcsec$ of the nominal target
position with a ``P'' label in the Notes column of Table
\ref{tab:continuum}.  These are our highest quality continuum data,
and for these sources alone the continuum data may be suitable for
deriving physical properties.  Due to confusion or poor data quality,
some sources were not detected in either or both continuum cross-scans
and in such cases we do not provide continuum intensities or angular
sizes.


\subsection{Multiple-velocity \hii\ Regions \label{sec:multvel}}
Of the 302 detections, 57 have spectra with multiple hydrogen RRL
components at different velocities.  In total, we detect the emission
from 369 RRL components (245 with one line, 47 sources with two lines,
and 10 sources with three lines).  As in \citep[][hereafter
  A15]{anderson15b} we hypothesize that one of these component is from
the discrete \hii\ region that we targeted and the other(s) are from
diffuse ionized gas along the line of sight.  To compute kinematic
distances, we must determine the correct RRL velocity for these
multiple-velocity \hii\ region spectra.

In A15 we derived a set of criteria that can be used to determine
which of the RRL components is from the discrete \hii\ region.  From
the data given here, we can use four of their criteria : (1) the
presence of a negative RRL velocity component, the association between
RRL velocities and those of either (2) molecular gas or (3) carbon
RRLs, and (4) an analysis of the electron temperature for each RRL
velocity component.  As in A15, we require that the results from all
four criteria agree with each other.  Below we discuss the four
criteria, and how we apply them to our data set.

For the first of the four criteria, A15 argue that diffuse ionized gas
is unlikely to be found in the outer Galaxy, at negative velocities in
the first Galactic quadrant, since the density of free electrons in
the outer Galaxy is low \citep[e.g.][]{taylor93}.  As in A15 we
therefore assume that for all first-quadrant multiple-velocity
\hii\ regions the negative velocity components are from the discrete
\hii\ regions.

Two of the four criteria rely on the detection of other spectral lines
within 10\,\kms\ of one of the \hnaa\ velocity components. A15 argue
  that molecular gas and carbon RRLs are more likely to be associated
  with discrete \hii\ regions than diffuse ionized gas.  For each
  multiple-velocity \hii\ region, we correlate the molecular
  velocities compiled by A14 with the \hnaa\ velocities measured here.
  We search for carbon lines by examining each spectrum for
    emission offset $\sim -150\,\kms$ from the hydrogen RRL velocity
    that is at least three times the rms, and correlate any such
    sources with the \hnaa\ velocities.  These correlations produce
  lists of velocity components with associated molecualar gas or
  carbon RRLs, which we assume are the discrete \hii region
  velocities.

Finally, A15 use the derived electron temperature values, \te, for
each RRL component to determine the discrete \hii\ region RRL
  components.  In local thermodynamic equilibrium (LTE), the electron
temperature can be computed from observable quantities:
\begin{equation}
T_e =  7103.3 \left(\frac{\nu}{\rm GHz}\right)^{1.1} \left[\frac{T_C}{T_L({\rm H}^+)}\right] \left[\frac{\Delta v({\rm H}^+)}{\rm km\,s^{-1}}\right]^{-1} \left[ 1 + \frac{n(^4{\rm He}^+)}{n({\rm H}^+)} \right]^{-1}\,,
\end{equation}
where $\nu$ is the observing frequency, $T_C/T_L$ is the
continuum-to-line intensity ratio, $\Delta v$ is the RRL FWHM line
width, and $n(^4{\rm He}^+)/n({\rm H}^+)$ is the helium ionic
abundance ratio by number, $y^+$.  Each multiple-velocity \hii\ region
has a single value for $T_C$, but multiple values for $T_L$ and
$\Delta v$, and therefore has a different \te\ for each RRL
component.  Assuming that only one of the detected lines is from a
discrete \hii\ region, the line-to-continuum intensity ratio is only
physically meaningful for this component.  Therefore, whereas the
electron temperature can be computed for all detected lines for each
multiple-velocity \hii\ region, the line from the \hii\ region should
have a reasonable value of \te\ while the others may not.  Based on
their analysis of single-velocity \hii\ regions from the HRDS, A15
define this ``reasonable'' range of \te\ values (in Kelvin) as a
function of Galactocentric radius, \rgal\ (kpc), for high quality
sources to be: $860 + 544 \rgal < \te < 2040 + 544 \rgal$.  A high
quality source is one with a peak continuum intensity of at least
100\,mK, a peak within $10\arcsec$ of the nominal centroid pointing,
and radio continuum emission that can be modeled with a single
Gaussian component.  A range for low-quality sources not meeting at
least one of the above criteria is defined by A15 to be: $\te < 6300 +
544 \rgal$.  Here, we compute a value of \te\ for each detected
hydrogen component using 8.7\,GHz for $\nu$, and assuming $y^+ =
0.07$ \citep{quireza06b}.  We use the ``reasonable'' range of values
above to identify which RRLs originate from discrete \hii\ regions.

\subsection{Distances}
We derive kinematic distances for 131 of the detected \hii\ regions.
Of these, 105 are outer-Galaxy regions ($\rgal > 8.5\,\kpc$) and 26
are inner-Galaxy regions.  The results of our kinematic distance
analysis, in addition to the numbers of regions known previously, are
summarized in Table~\ref{tab:distance_summary}.

Kinematic distances use a model for Galactic rotation to give
distances as a function of observed velocity for a given line of
sight.  We derive all kinematic distance parameters using the
\citet{brand93} rotation curve.  Kinematic distances are prone to
large uncertainties in certain parts of the Galaxy.  As in previous
work, we estimate kinematic distance uncertainties by adding in
quadrature the uncertainties associated with the rotation curve
choice, streaming motions of 7\,\kms, and changes to the Solar
circular rotation speed.  Such uncertainties were first computed by
\citet{anderson12c}, and expanded to the entire Galaxy by A14.  We use
the latter analysis here.

Kinematic distances are possible for 185 nebulae.  We do not
compute kinematic distances for sources within $10\degree$ of the
Galactic center or within $20\degree$ of the Galactic anti-center
because such distances would be uncertain by $\gtrsim 50\%$ (A14).  We
also exclude \hii\ regions with multiple RRL velocities for which the
source velocity is unknown, as they have at least two possible
kinematic distances.  We exclude sources in the first and fourth
Galactic quadrants for which the absolute value of the tangent point
velocity is less than 10\,\kms.  Finally, we do not provide distances of
outer Galaxy regions whose distance uncertainties are $>50\%$ that of
their kinematic distances.

Sources in the inner Galaxy suffer from the well-known kinematic
distance ambiguity (KDA): inner Galaxy \hii\ regions have two possible
Heliocentric distances (called ``near'' and ``far'') for each measured
velocity.  The KDA does not exist for the 105 Outer Galaxy
\hii\ regions in our sample (including the 34 first-quadrant sources
with negative RRL velocities and the four fourth quadrant sources with
positive velocities).  For all regions, there is no ambiguity in the
calculation of Galactocentric distances.

There are 80 inner-Galaxy \hii\ regions in our sample for which
kinematic distances are possible.  We provide a kinematic distance
ambiguity resolution (KDAR) for 26 of these (33\%).  Four of these are
Sharpless regions, which we assume lie at their near distances since
they are optically visible (note, however, the extreme distance to S83
discussed later).  We analyze the remaining 76 first-quadrant nebulae
using the \hi\,E/A method \citep{anderson09a, anderson12c}, employing
VGPS 21\,cm data.  The \hi\,E/A method uses the fact that \hi\ between
the observer and the \hii\ region is detected in absorption when the
brightness temperature of the \hii\ region plasma at 21\,cm is greater
than that of the foreground \hi.  \hi\ beyond the \hii\ region will be
seen in emission.  
This method becomes unreliable if the source velocity is near the
tangent point, and therefore we give the six nebulae that have LSR
velocities within 10\,\kms\ of the tangent point the tangent point
distance.  We are only able to provide a KDAR for 16 of the remaining
70 inner-Galaxy nebulae.  To determine distances to the other 54
  nebulae we would need more sensitive \hi\ observations.

We give the kinematic distance parameters for all 131 \hii\ regions
with kinematic distances in Table~\ref{tab:distances}, which lists
the source name, the LSR velocity, the near, far, and tangent point
distances, the Galactocentric radius, the tangent point velocity, the
KDAR (``N'' = near distance, ``F'' = far distance, ``T'' = tangent
point distance, ``O'' = outer Galaxy), the
Heliocentric distance (with uncertainties), and
the vertical distance, $z$, from the Galactic mid-plane.

In addition to the kinematic distances described above, there are a
number of regions that based on their \lbv\ coordinates are physically
near to the Galactic center.  We observed four single-velocity regions
that are associated with Sgr~B2, and four that are associated with
Sgr~E.  These two star forming complexes are physically near to the
Galactic center \citep[$\sim 100$\,pc][]{reid09c, liszt92}.  While we
do not provide distance parameters for these sources in
Table~\ref{tab:distances}, in later plots we do use a Heliocentric
distance of 8.5\,kpc\ and a Galactocentric radius of 0.1\,\kpc\ for
these 8 regions.


\section{Discussion\label{sec:properties}}

\subsection{Sharpless Regions\label{sec:sharpless}}
\begin{figure} \centering
\includegraphics[width=6in]{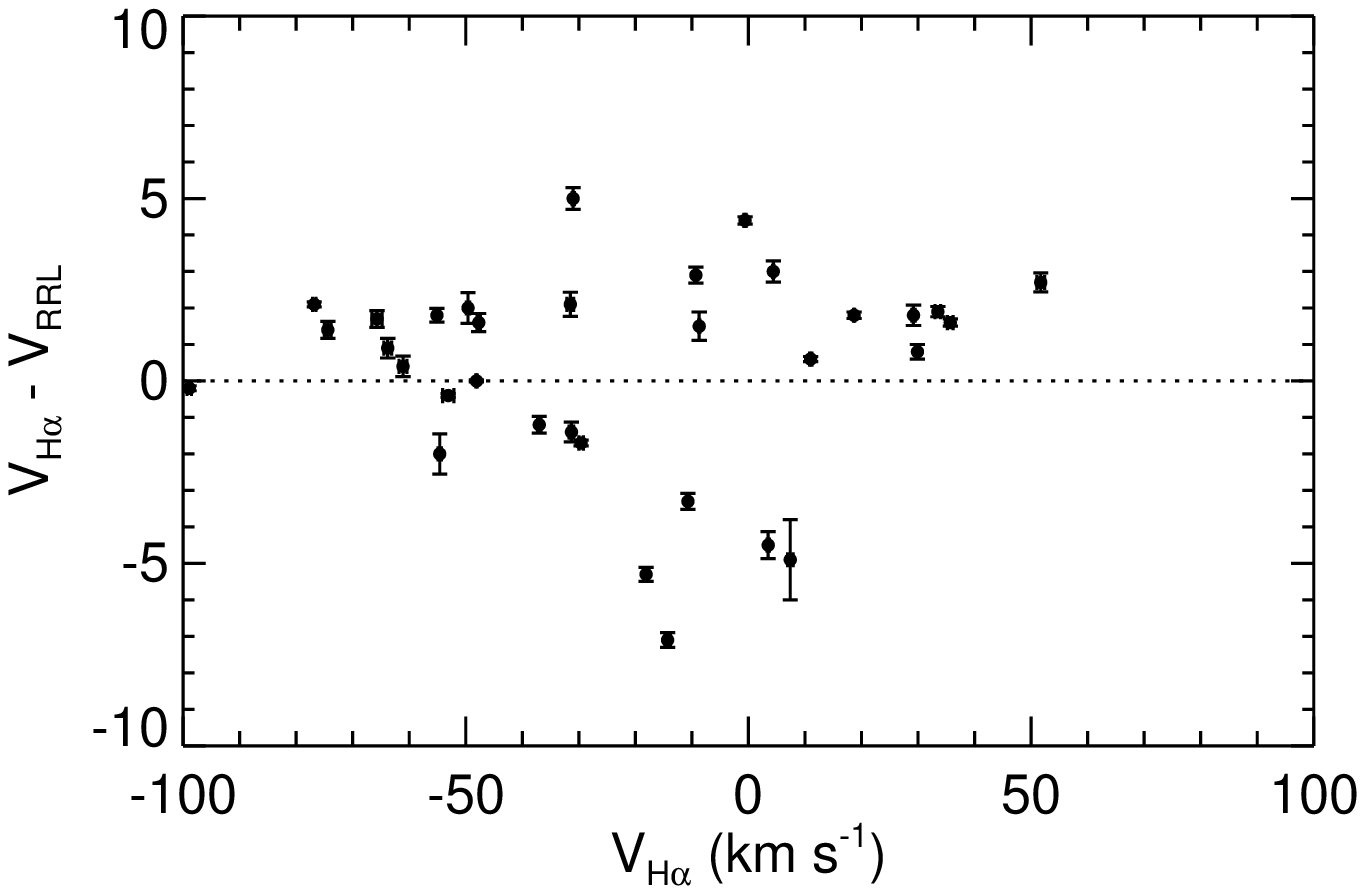}
\includegraphics[width=6in]{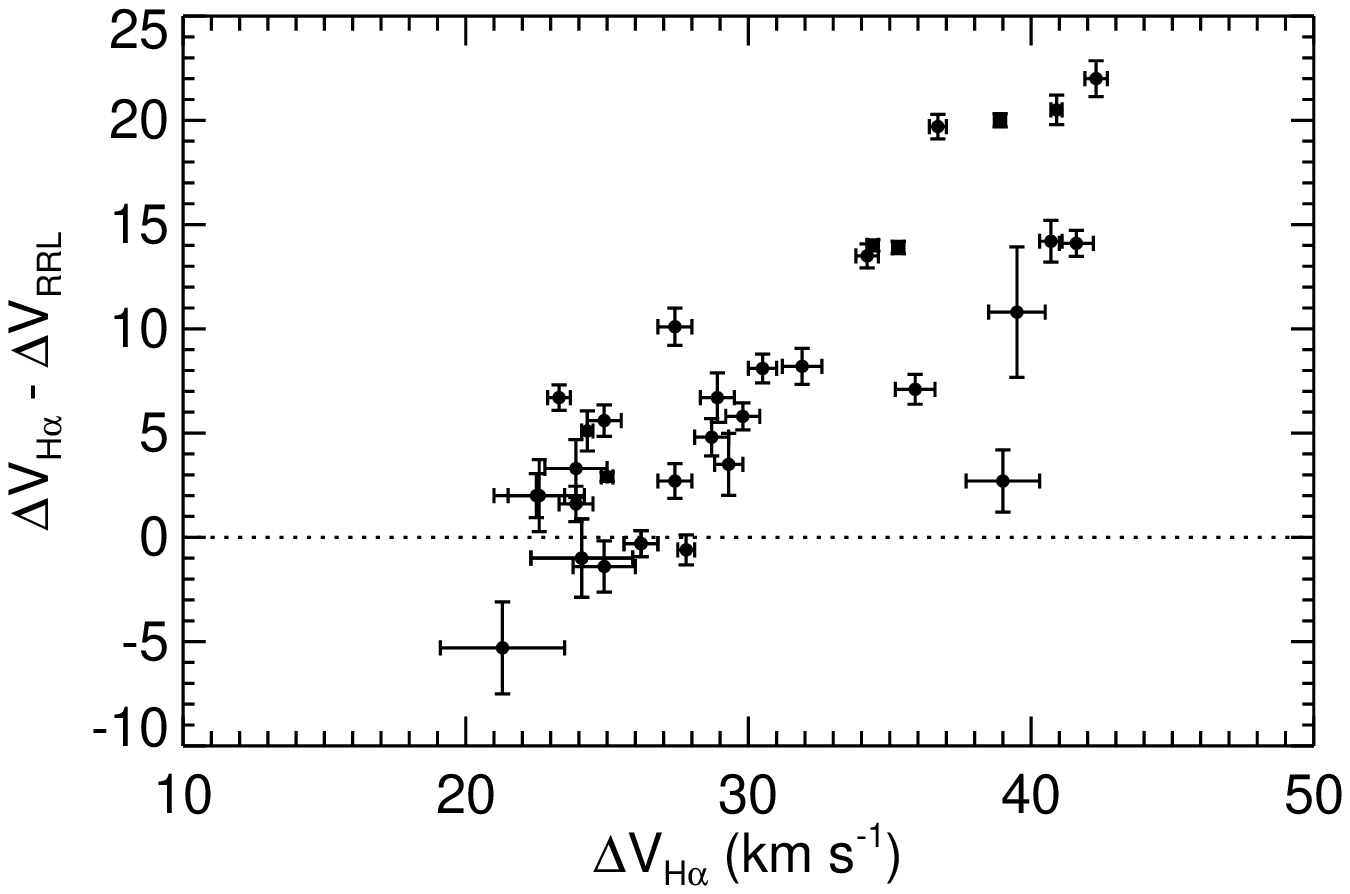}
\caption{Comparison of H$\alpha$ \citep[from][]{fich90} and RRL
  velocities (top) and FWHM line widths (bottom) for observed
  Sharpless \hii\ regions.  The velocity differences are small
  compared to the $\sim15$\,\kms\ H$\alpha$ velocity resolution and
  there is no systematic offset.  The line widths, however, are
  significantly broader for the H$\alpha$ observations, in agreement
  with the finding of \citet{fich90}.
\label{fig:sharpless}}
\end{figure}

The Sharpless regions we detected have velocities similar to those
observed in H$\alpha$ by \citet{fich90}, although the FWHM line widths
are discrepant.  We show these comparisons for the \hii\ regions with
one velocity component in Figure~\ref{fig:sharpless}.  By comparing
with the H109$\alpha$ measurements of \citet{lockman89},
\citet{fich90} also found very good agreement between H$\alpha$ and
RRL velocities, but large discrepancies between the line width
measurements.  They note a mean H$\alpha$ and RRL velocity difference
of $0.82\pm3.46$\,\kms, whereas we find $0.38\pm2.73\,\kms$ (quoted
uncertainties here are the dispersion).  The mean absolute velocity
difference ($<|v_{\rm RRL} - v_{\rm H\alpha}|>$) is
$2.02\pm1.41\,\kms$.  The H$\alpha$ line widths are systematically
greater than those we measure for RRLs, in agreement with the results
of \citet{fich90}.  The average line width ratio $\Delta V_{\rm RRL} /
\Delta V_{\rm H\alpha}$ is $0.82\pm0.17$, whereas \citet{fich90} found
$0.83\pm 0.18$.  Therefore, the line width discrepancy with H$\alpha$
is the same, regardless of the observed radio frequency. We
hypothesize that the H$\alpha$ observations are broadened by the
relatively low spectral resolution of $\sim15\,\kms$.  We can
  spectrally deconvolve the H$\alpha$ linewidths using $\Delta V_{\rm
    true}^2 = \Delta V_{H\alpha}^2 - \Delta V_{\rm Res.}^2$, where the instrumental spectral resolution 
  $V_{\rm Res.} = 15\,\kms$.  Doing so, we find that the average line
  width ratio $\Delta V_{\rm RRL} / \Delta V_{\rm H\alpha}$ is
  $0.90\pm0.24$.  The spectral resolution can therefore only explain
  part of the linewidth difference.

\begin{figure} \centering
\includegraphics[width=6.5in]{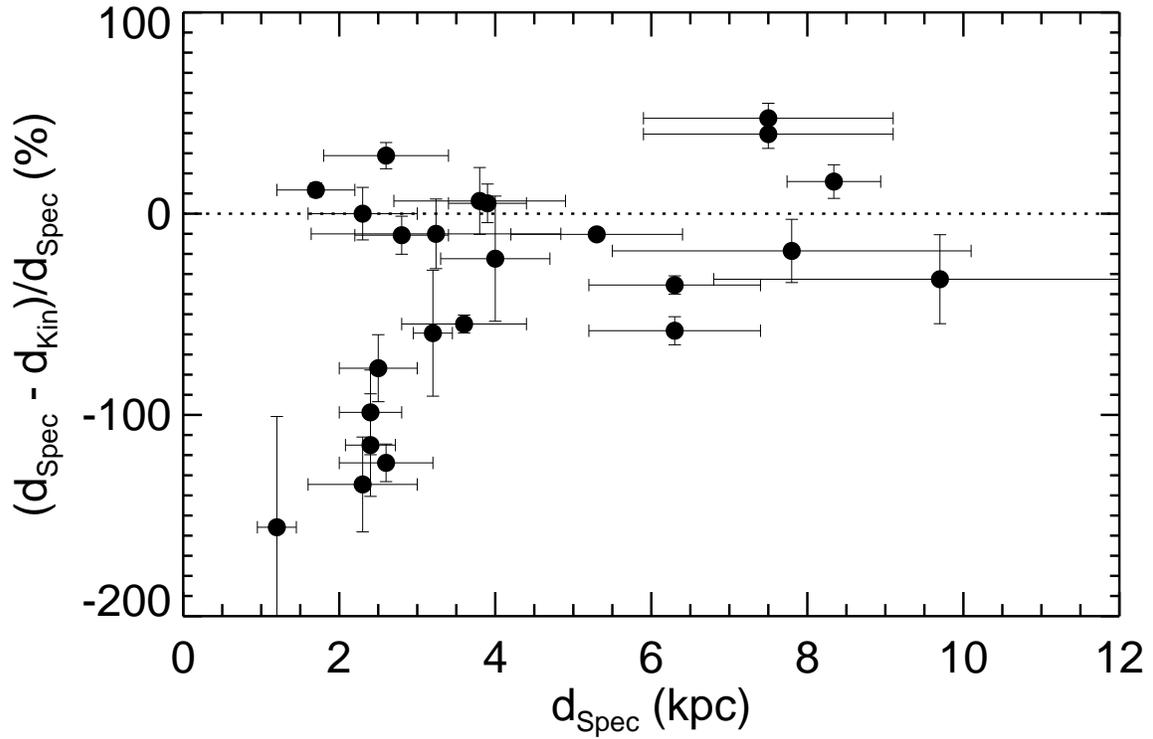}
\caption{Comparison of the percentage difference between
  spectrophotometric distances and kinematic distances derived here.
  The differences are large on average, but are generally less than
  100\% discrepant. The largest discrepancies are for sources
    with small spectrophotometric distances.
\label{fig:sharpless_distances}}
\end{figure}

There are 24 Sharpless \hii\ regions that have both spectrophotometric
distances and kinematics distances derived here. We give the
spectrophotometric and kinematic distances for these regions in
Table~\ref{tab:sharpless} and compare the distances in
Figure~\ref{fig:sharpless_distances}.  We take the spectrophotometric
distance values and uncertainties from \citet{russeil07} if possible,
or otherwise from \citet{russeil03}.  While there are more recent
spectrophotometric distances for some regions, using distances from
these two papers reduces uncertainties caused by different
methodologies.  The two distances are correlated but there is
considerable scatter, especially for sources with small
spectrophotometric distances.  The mean difference ($<\!\!d_{\rm Kin}
- d_{\rm Spec}\!\!>$) between kinematic and spectrophotometric
distances is $1.0\pm2.0$\,kpc.  As a percentage of the
spectrophotometric distances, the differences are on average
$47\pm90$\% discrepant.  The average absolute difference
($<\!\!|d_{\rm kin} - d_{\rm spec}|\!\!>$) in kinematic and
spectrophotometric distances is $1.8\pm 1.3\,\kpc$.  The agreement is
improved if the \citet{reid14} rotation curve is used.

The largest discrepancies between distances are for \hii\ regions with
small spectrophotometric distances in the zone $136\degree > \ell >
105\degree$.  For example, S135, S164, S170, S175, and S193 all have
spectrophotometric distances $<2.5$\,kpc, but kinematic distances from
the \citet{brand93} rotation curve $\ge 100\%$ discrepant.  The
\lbv\ locii of these regions suggests that they lie in the Perseus
spiral arm where perhaps non-circular motions are leading to
erroneously large kinematic distances \citep{choi14}.  This
explanation assumes that the spectrophotometric distances are more
accurate than the kinematic distances, which has not been proven to be
the case.  We do note that for S170 the maser parallax distance of
3.34\,\kpc \citep{choi14} is in better agreement with the
spectrophotometric distance of 2.6\,\kpc compared with our kinematic
distance of 5.8\,\kpc.  \citet{choi14} list kinematic distances for
this source of 3.0\,\kpc and 4.0\,\kpc\ for different rotation curve
models based on their maser parallax distances, which implies that the
\citet{brand93} curve dramatically overestimates kinematic distances
in this direction.

\subsection{Galactic Distribution\label{sec:galactic_structure}}
\begin{figure} \centering
\includegraphics[width=6.5in]{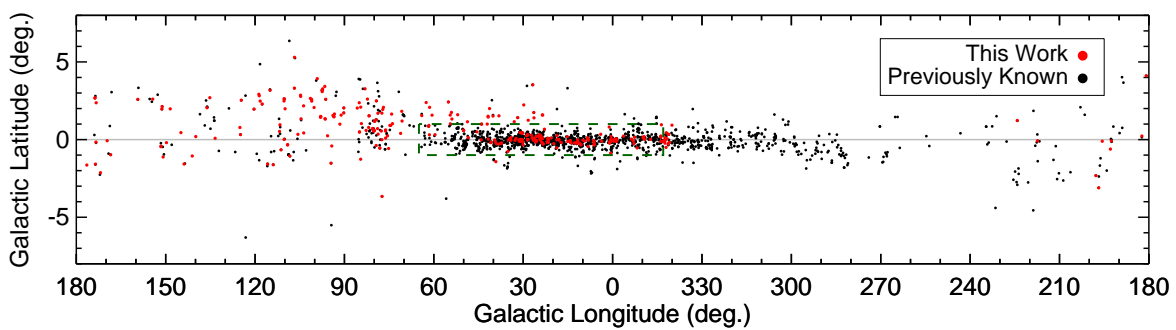}
\caption{Galactic distribution of the detected \hii\ regions (red) and
  those known previously (black).  The previously known sample is from
  A14 and includes regions from the original HRDS, whose survey zone,
  $65\degree > \ell > -17\degree$; $\absb \le 0\degree$, is indicated
  by the dashed box. Outside the range of the original HRDS the new
  detections are especially numerous compared to the sample of
  previously known \hii\ regions, with 170 new detections over
    the survey area versus 214 regions known previously.  The gray
  horizontal line indicates $b=0\degree$.  The new detections have a
  broader latitude distribution in the first quadrant than those
  previously known. \label{fig:lb}}
\end{figure}

Here, we have added significant numbers of \hii\ regions to the
Galactic census, especially in the sky zones not covered by the
original HRDS.  In Figure~\ref{fig:lb} we show the distribution of
regions detected here and of all \hii\ regions known previously.  The
sample of previously known regions all have ionized gas velocities
(RRL or H$\alpha$), as compiled in the WISE catalog (A14).  The sample
here adds considerably to the \hii\ region census in the portion of
the first Galactic quadrant not covered by the original HRDS,
$90\degree > \ell > 65\degree$, with 42 new detections versus 52
regions known previously.  We also add 105 outer Galaxy regions
(mostly in the second quadrant) versus 160 known previously
(Table~\ref{tab:distance_summary}).

\begin{figure} \centering
\includegraphics[width=5.25in]{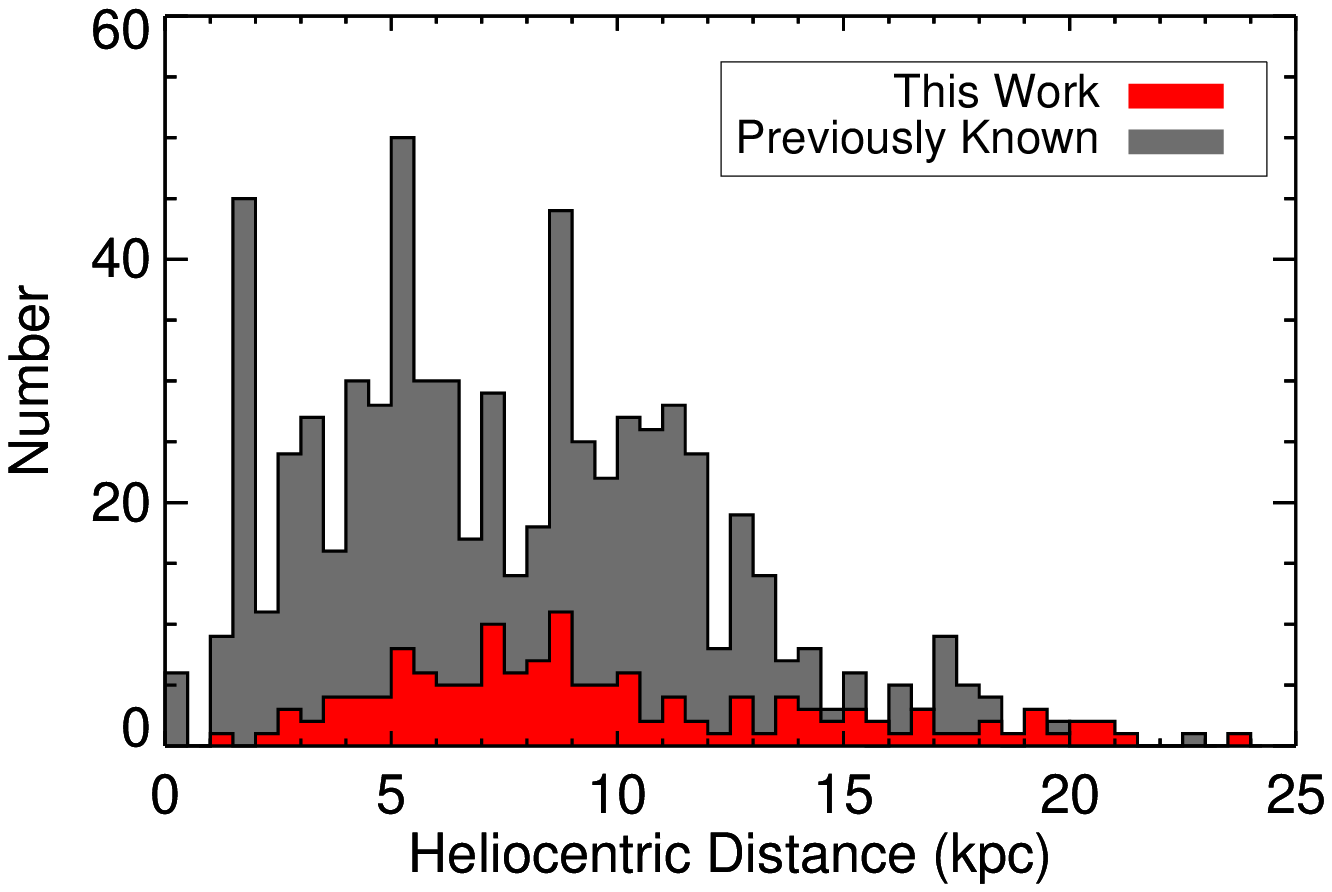}
\includegraphics[width=5.25in]{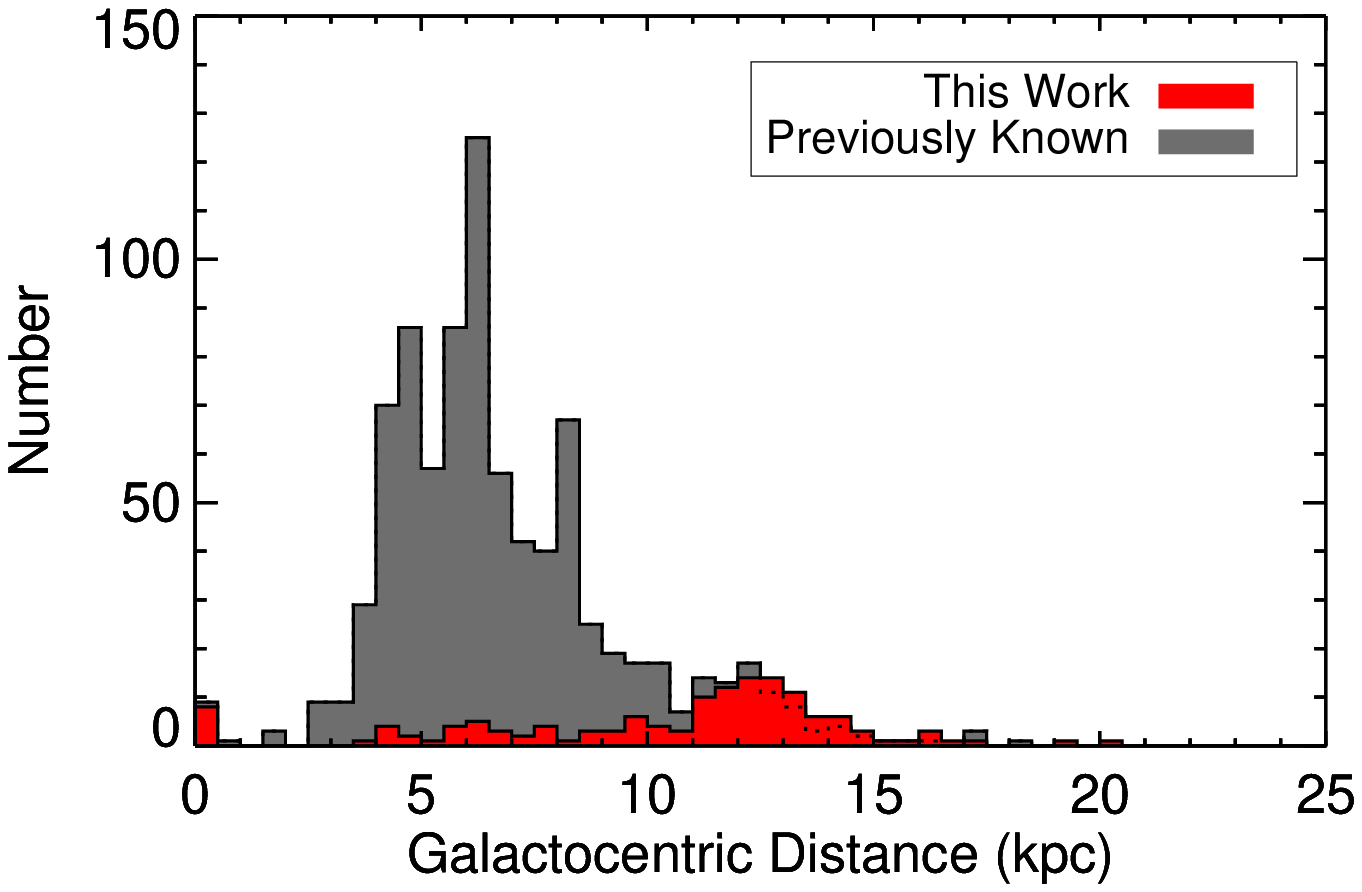}
\caption{Heliocentric (top) and Galactocentric (bottom) \hii\ region
  distance distributions for the new detections (red) and those
  previously known (gray).  The previously known sample contains all
  \hii\ regions with a single measured ionized gas velocity component
  (H$\alpha$ or RRLs) compiled in the WISE catalog of A14, incuding
  regions from the original HRDS.  Here we show all nebulae located in
  the zone ($225\degree > \ell > -20\degree$).  Neither sample shown
  here contains sources with multiple velocity components.  The newly
  detected \hii\ regions are more distant on average than those known
  previously, in terms of both Heliocentric and also Galactocentric
  distances.  The average Heliocentric and Galactocentric
    distances for the new detections are 9.8\,\kpc and
    11.0\,\kpc, respectively, while they are 7.4\,\kpc
    and 6.8\,\kpc for the previously known sample.}
\label{fig:distances}
\end{figure}

As expected, the regions in the current sample are more distant on
average than the population known previously.  To illustrate this
point, in Figure~\ref{fig:distances} we show the distributions of
Heliocentric and Galactocentric distances for the current census of
Galactic \hii\ regions.  The average Heliocentric and Galactocentric
distances for the new detections are 9.8\,\kpc and 11.0\,\kpc,
respectively, while they are 7.4\,\kpc and 6.8\,\kpc for the
previously known sample compiled by A14.  Here we provide kinematic
distances for 60 newly-found first-quadrant regions.  The average
Heliocentric distance to these 60 regions is 15.2\,kpc, whereas it is
10.1\,\kpc\ for the first-quadrant regions in the original HRDS
\citep{anderson12c} and 8.4\,\kpc\ for the first-quadrant regions
known prior to the HRDS \citep{anderson09b}.  The average
Galactocentric radius for the 60 first-quadrant sources is 9.4\,\kpc.
Of the 60 regions, 34 have negative RRL velocities and are therefore
in the outer Galaxy.  In the second quadrant, the average distance is
7.2\,kpc, whereas it was 5.7\,kpc for regions known previously.  We
attribute these differences to the fact that we explicitly searched
for distant regions following the warp in the first quadrant, and
observed fainter sources than were known previously in the second
quadrant.

\subsection{Outer Scutum Centaurus Arm}
We detected the RRL emission from 47 \hii\ regions in the \lb\ zone of
the OSC searched here: $65\degree > \ell > 25\degree; b \ge 1\degree$.
Of these 47 \hii\ regions, 5 have LSR velocities within 15\,\kms\ of
the OSC \lv\ locus defined by in \citet{dame11}: $v = -1.6 \times
\ell$.  Included in this tally is S83, whose RRL velocity was already
known to be consistent with the OSC \citep[e.g.,][where it is called
  ``G55.11+2.4'']{balser11}.  We also detect the RRL emission from an
additional region not in the OSC \lb\ zone whose longitude and
velocity are nevertheless consistent with membership in the OSC,
G039.183$-$01.422 (see below), bringing the total number of
\hii\ regions identified here that appear to be in the OSC
to 6: G033.008+01.151\hide{DB015}, G039.183$-$01.422\hide{FQ009},
G041.755+01.451\hide{DB028}, G041.804+01.503\hide{DB029},
G054.094+01.749\hide{DB067}, and S83.  The regions
G041.755+01.451\hide{DB028} and G041.804+01.503\hide{DB029} have small
angular separations and have RRL velocities within $\sim2$\,\kms\ of
each other.  They are therefore presumably part of the same complex.

We give the parameters of the OSC regions in Table~\ref{tab:osc},
where we list Galactocentric and Heliocentric distances according to
the \citet{brand93} (with $\Theta_0 = 220$\,\kms\ and $R_0 =
8.5$\,kpc) and \citet{reid14} rotation curves (with $\Theta_0 =
240$\,\kms\ and $R_0 = 8.34$\,kpc).  The \citet{reid14} curve gives
distances $\sim10\%$ smaller for the OSC sources, on average $\sim
2$\,kpc for distances of $\sim20\,\kpc$, due to the larger rotational
speeds of the model.

The source G039.183$-$01.422 has an LSR velocity consistent with being
in the OSC, and yet according to its kinematic distance it lies nearly
500\,pc below the Galactic plane.  At $\ell = 39\degree$, the \hi\ in
the OSC is centered near $b = 3\degree$ \citep[c.f.][]{dame11}
or $\sim600$\,pc above the Galactic mid-plane at a Heliocentric
distance of $\sim20$\,\kpc.  There is only a small amount of OSC
\hi\ emission seen at the \lb\ position of G039.183$-$01.422
\citep[see][]{dame11}.  The implied separation of G039.183$-$01.422
from the OSC arm centroid therefore corresponds to a vertical offset
of $>1$\,\kpc.  It is unlikely that the high-mass stars capable of
producing \hii\ regions would have strayed this far from their
birthplaces. At the nominal lower velocity for a star to be
  called a ``runaway,'' 30\,\kms, it would take over 30\,Myr to travel
  1\,kpc.  This neglects motion in the plane of the sky.  Of course,
  if it is indeed a runaway star, its velocity cannot be used for
  kinematic distances.  It is possible that G039.183$-$01.422 is a
planetary nebula, but we regard this as unlikely due to the detection
of an \water\ maser at $-53$\,\kms\ \citep{sunada07} as well as NH$_3$
(W.~Armentrout, 2015, in prep.).

\subsection{Distant Outer Galaxy Sources Toward $\ell = 150\degree$}
Three second-quadrant sources have Galactocentric radii from the
\citet{brand93} rotation curve that are greater than 19\,\kpc:
G149.746$-$00.199, G151.561$-$00.425, and G151.626$-$00.456.  Based on
their WISE emission, the latter two sources appear to be in the same
complex as S209, which has a RRL velocity of
$-51.2$\,\kms\ \citep{balser11}.  Properties of these nebulae are
included in Table~\ref{tab:osc}.  The \citet{reid14} curve gives
Galactocentric distances lower by $3-5\,\kpc$ ($\sim20\%$;
Table~\ref{tab:osc}).  Regardless of their exact distances, these
sources have the largest Galactocentric radii of any \hii\ regions
discovered and are likely located in the extreme outer Galaxy
\citep[EOG][]{kobayashi08, yasui08}.  With low metallicity and low
densities, the EOG is a laboratory for how stars may have formed in
the early evolutionary stages of our Milky Way.

The three distant sources may be part of the OSC extension into the
second Galactic quadrant that was identified by \citet{sun15}, whose
analysis shows \hi\ emission at \lv = ($\sim150\degree$,
$\sim-70$\,\kms).  While the \hii\ region velocities are in rough
agreement with this \lv\ locus, the brightest \hi\ emission near
$\ell\simeq 150\degree$ is at $b\simeq -3\degree$ whereas our sources
are much closer to $b=0\degree$.  It is therefore not clear at present
if these three \hii\ regions (and S209) lie in the OSC.  These distant
\hii\ regions are not spatially coincident with any of the molecular
clouds detected by \citet{digel94}, which have Galactic longitudes of
$130\degree-150\degree$ and Galactocentric radii $\gtrsim17\kpc$.

\subsection{Nuclear Disk (Sgr E)}


We detected RRL emission from 5 \hii\ regions near $\ell = 359\degree$
that have negative velocity components $\lsim-200$\,\kms:
G358.600$-$00.057\hide{E33}, G358.643$-$00.034\hide{E34},
G358.684$-$00.116\hide{E55}, G358.802$-$00.011\hide{E49}, and
G358.946+00.004.  These nebulae are part of Sgr~E, a well-known
complex of \hii\ regions near the Galactic center \citep{liszt92} that
contains at least 30 separate sources identified from their
  spectral indices \citep{gray93}.  Few of these 30 regions have
  ionized gas spectroscopic measurements.  Sgr~E is thought to be at the outer
boundary of the nuclear disk, which implies a distance from the
Galactic center of $\sim200$\,pc.  It has no counterpart at positive
longitudes.  Sgr~E was previously observed in RRL emission by
\citet{cram96}, and they detected 6 \hii\ regions, including four of
the regions detected here.  The region G358.946+00.004, which was not
detected by \citet{cram96}, has multiple RRL velocities and therefore
its membership in Sgr~E is less secure.

\begin{figure}
\includegraphics[width=6.5in]{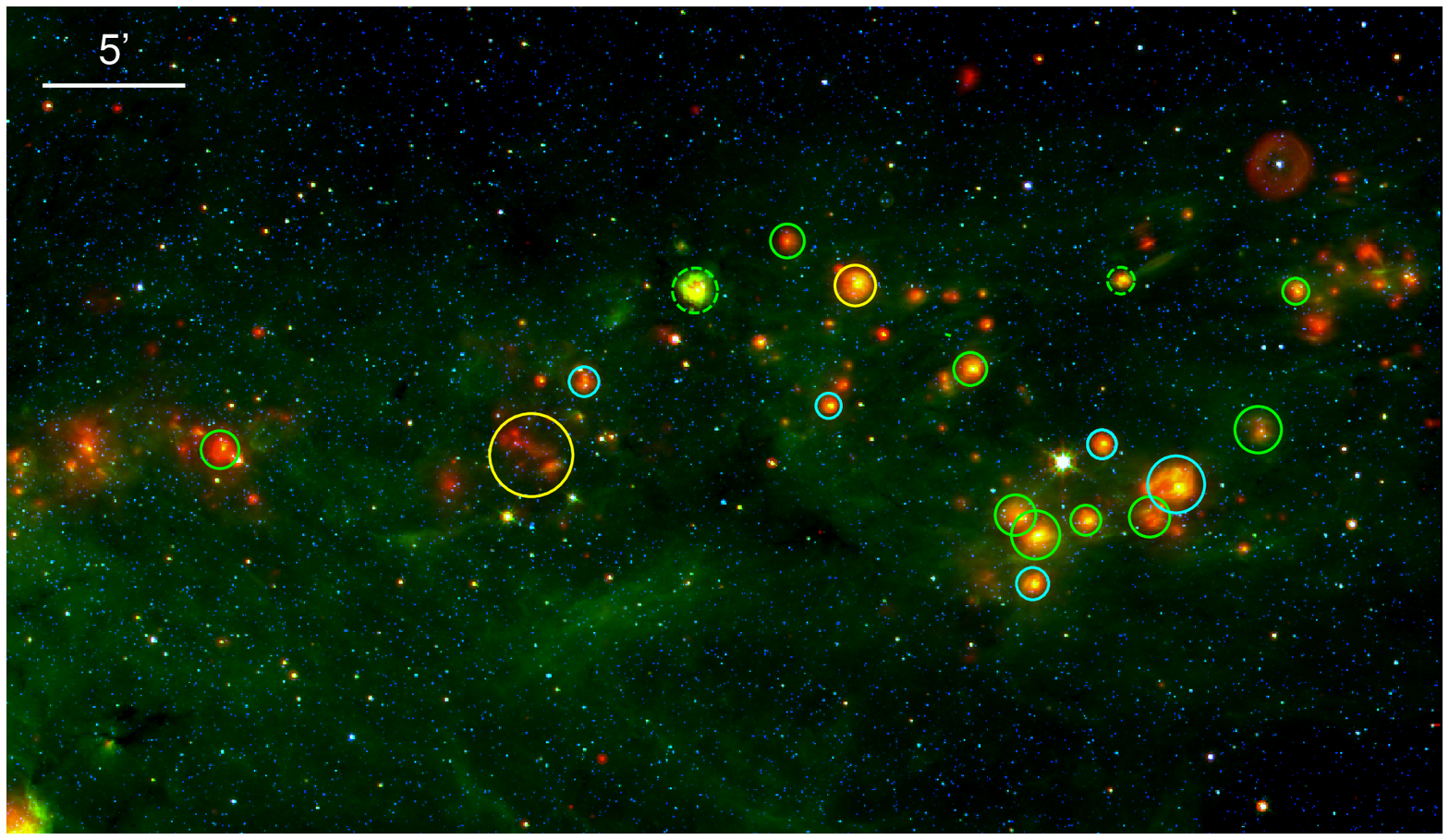}
\caption{{\it Spitzer} image of nuclear disk \hii\ regions in Sgr~E.
  The field is $50\arcmin\times30\arcmin$, centered at \lb =
  (359.863\degree, $-$0.020\degree). The 24\,\micron\ MIPSGAL data are
  red, the 8.0\,\micron\ GLIMPSE data are in green, and the
  3.6\,\micron\ GLIMPSE data are in blue.  Circles approximate the
  \hii\ region infrared sizes and are color-coded such that cyan 
    circles enclose \hii\ regions from this work, green are from the
  original HRDS (A11), and yellow are from \citet{lockman96}.  The two
  dashed circles surround regions that have velocities inconsistent
  with that of the nuclear disk.  Other sources evident in the field
  lack ionized gas spectroscopic velocities.  The nuclear disk sources
  have ratios of 24\,\micron\ to 8.0\,\micron\ fluxes about three
    times that of \hii\ regions found in the rest of the
  Galaxy.}
\label{fig:nuclear}
\end{figure}

On the basis of their Galactic
location and LSR velocities near $-200\,\kms$, Sgr~E now has 16
\hii\ regions with measured RRL emission, including 9 regions from the
original HRDS \citep[one of which, G358.720+0.011, was also observed
  by][]{cram96}.  There are two additional \hii\ regions near $\lb =
(359\degree, 0\degree)$ with velocities near 0\,\kms\ whose membership
in Sgr~E is unlikely: G358.881+00.057; $-7.0$\,\kms (detected here)
and G358.633+00.062; $14.0$\,\kms (detected by A11).  We show the
infrared emission from the Sgr~E region in Figure~\ref{fig:nuclear},
which has 24\,\micron\ {\it Spitzer} MIPSGAL data in red,
8.0\,\micron\ {\it Spitzer} GLIMPSE data in green, and
3.6\,\micron\ {\it Spitzer} GLIMPSE data in blue.

The morphology of the Sgr~E \hii\ nebulae is unusual in that their
8.0\,\micron\ emission is weak compared to their
24\,\micron\ emission.  While there is faint 8.0\,\micron\ emission
surrounding each source, the strongest 8.0\,\micron\ emission is found
toward the center of the \hii\ regions.  This unusual infrared
morphology is not seen outside of the Galactic center. Furthermore,
for the two regions whose velocities are {\it not} consistent with
Sgr~E, the 8.0\,\micron\ emission {\it is} strong, and found
surrounding the regions.

The integrated 8.0\,\micron\ to 24\,\micron\ ratios are on average
$\sim 3$ times less than the Galactic average for \hii\ regions from
\citet{anderson12a}.  Why the 8.0\,\micron\ emission in the Sgr~E
\hii\ regions is so weak is unknown.  The 8.0\,\micron\ emission may
be attenuated by intervening material, the emission process that
produces the 8.0\,\micron\ emission may be absent for Sgr~E, or the
photodissociation regions surrounding the Sgr~E regions may not be
present.




\section{Summary\label{sec:summary}}
We report the Green Bank Telescope (GBT) detection of 302
\hii\ regions in radio recombination line and radio continuum
emission.  These regions were identified as \hii\ region candidates
from their mid-infrared and radio continuum emission in the WISE
Catalog of Galactic \hii\ Regions \citep{anderson14}, and have
estimated peak flux densities $>30$\,mJy at X-band.  The regions are
distributed throughout the entire sky visible from the GBT
($220\degree > \ell > -20\degree$ at $b = 0\degree$).  The largest
number of detections is in areas of the sky not studied in our
previous \hii\ Region Discovery Survey (HRDS), namely $\absb >
1\degree$, $\ell < -17\degree$, and $\ell > 65\degree$.  In these
zones the present survey has roughly doubled the number of known
\hii\ regions, bringing the total number of \hii\ regions detected in
our RRL surveys to nearly 800.  Here, we provide kinematic distances
for 131 regions.

Many of the new detections are extremely distant.  We detect 5
\hii\ regions that have \lbv coordinates consistent with the Outer
Scutum-Centaurus Arm (OSC), and one additional region with a velocity
consistent with the OSC but at a negative Galactic latitude
(G039.183$-$01.422; $-54.9\,\kms$).  The OSC is the most distant
molecular spiral arm of the Milky Way and these \hii\ regions are the
most distant ever discovered.  The OSC regions' Heliocentric distances
are $>22$\,kpc, and their Galactocentric distances are $>16$\,kpc.
The OSC regions were missed in previous surveys of the Galactic plane
because the OSC is found at Galactic latitudes $>1\degree$ in the
first Galactic quadrant. We detect an additional three \hii\ regions
near $\ell \simeq 150\degree$ whose LSR velocities imply
Galactocentric distances $>19$\,kpc.  These nebulae may be part of an
extension of the OSC, and they may represent the limit to Galactic
massive star formation.


\appendix
\section{Web Sites}
\begin{figure} \centering
\includegraphics[height=1.3in]{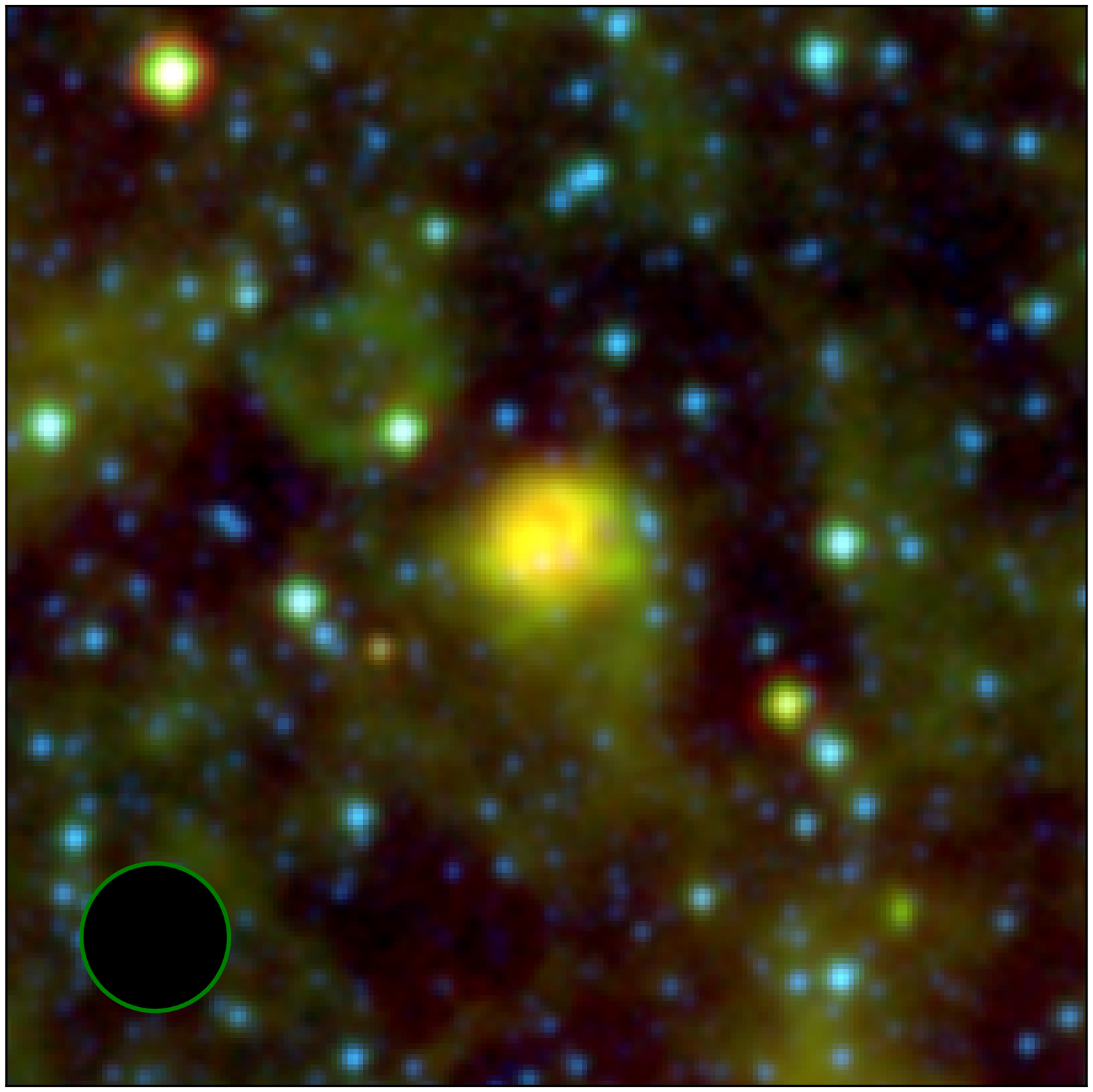}
\includegraphics[height=1.5in]{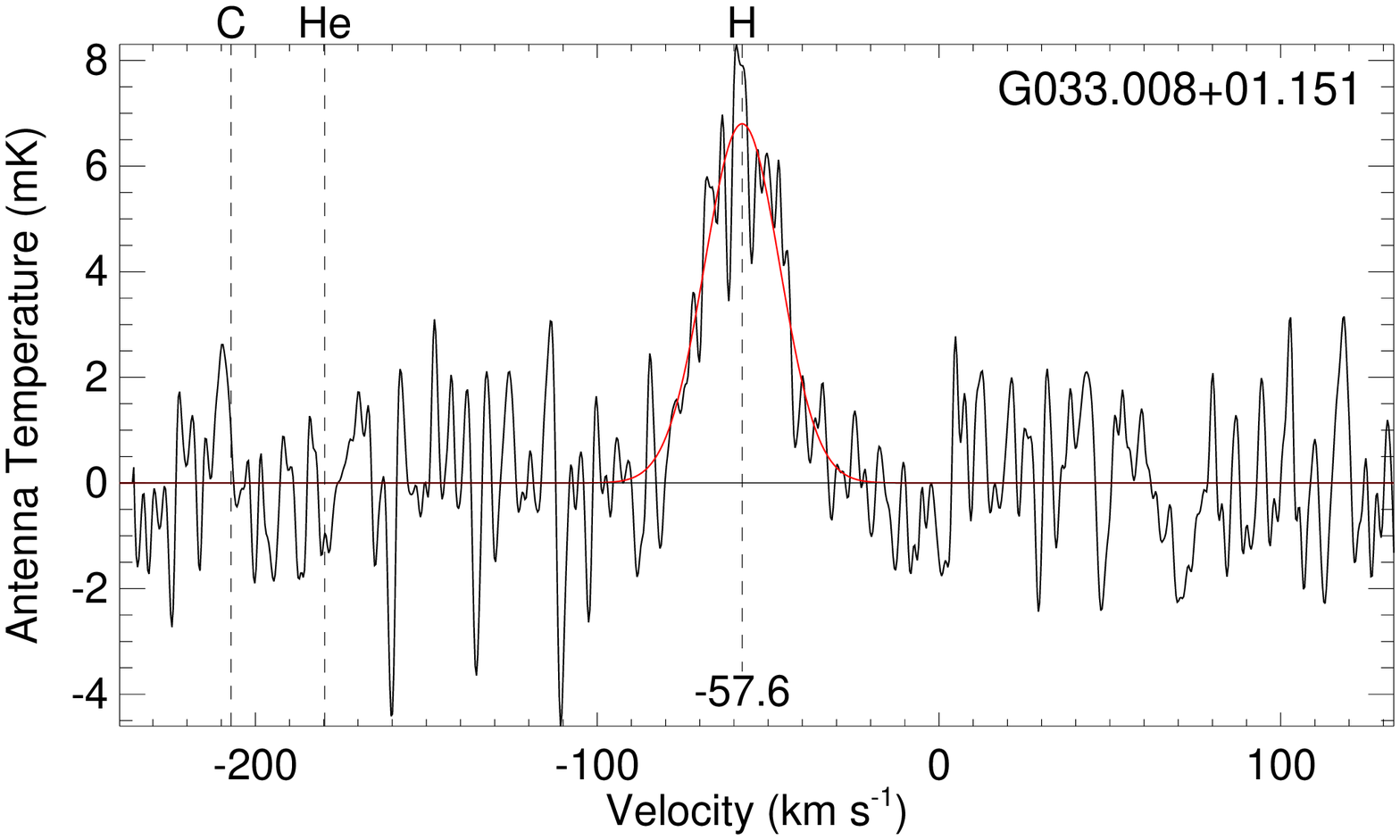}
\includegraphics[height=1.5in]{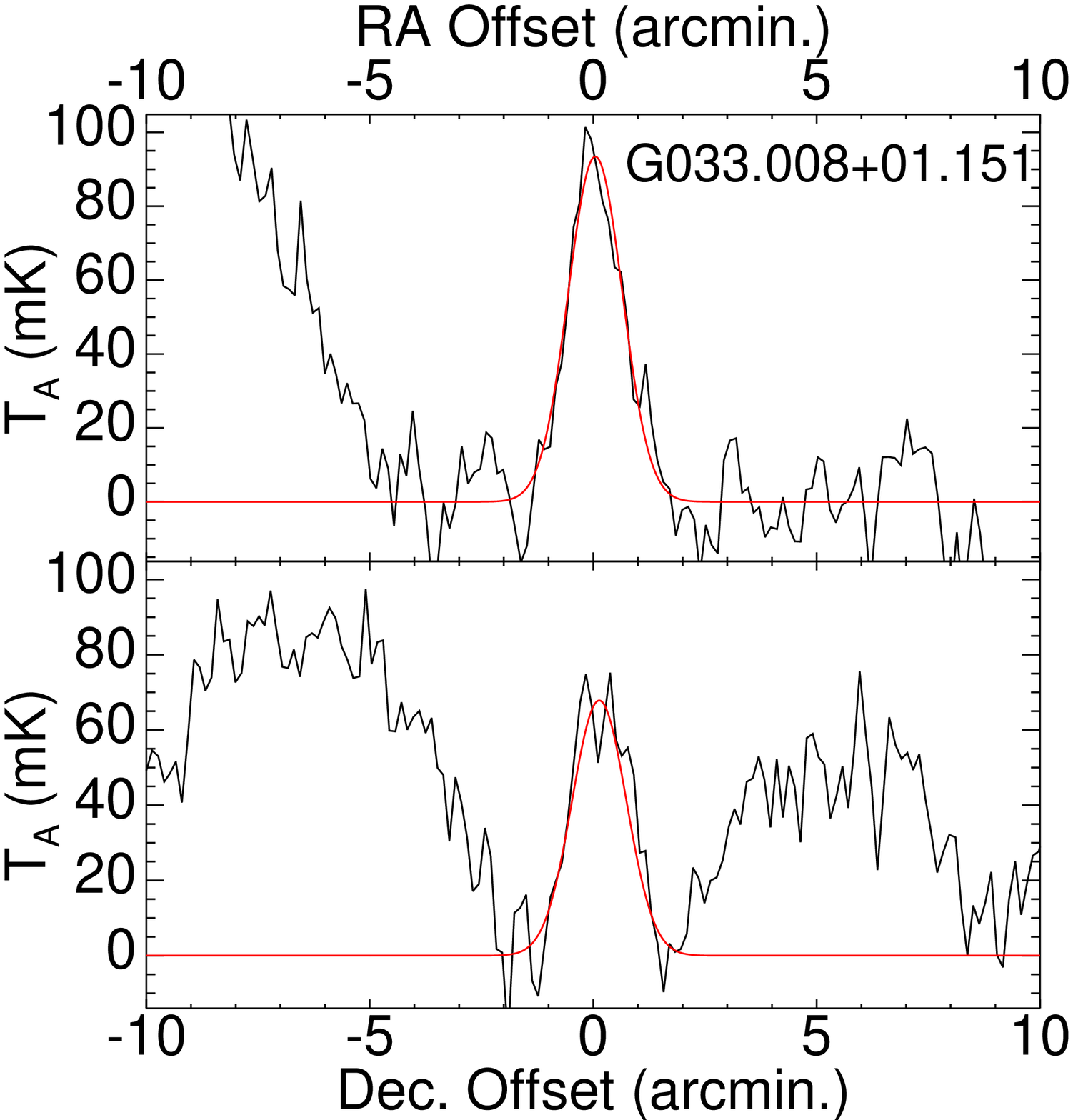}\\
\vskip 0.1in
\includegraphics[height=1.3in]{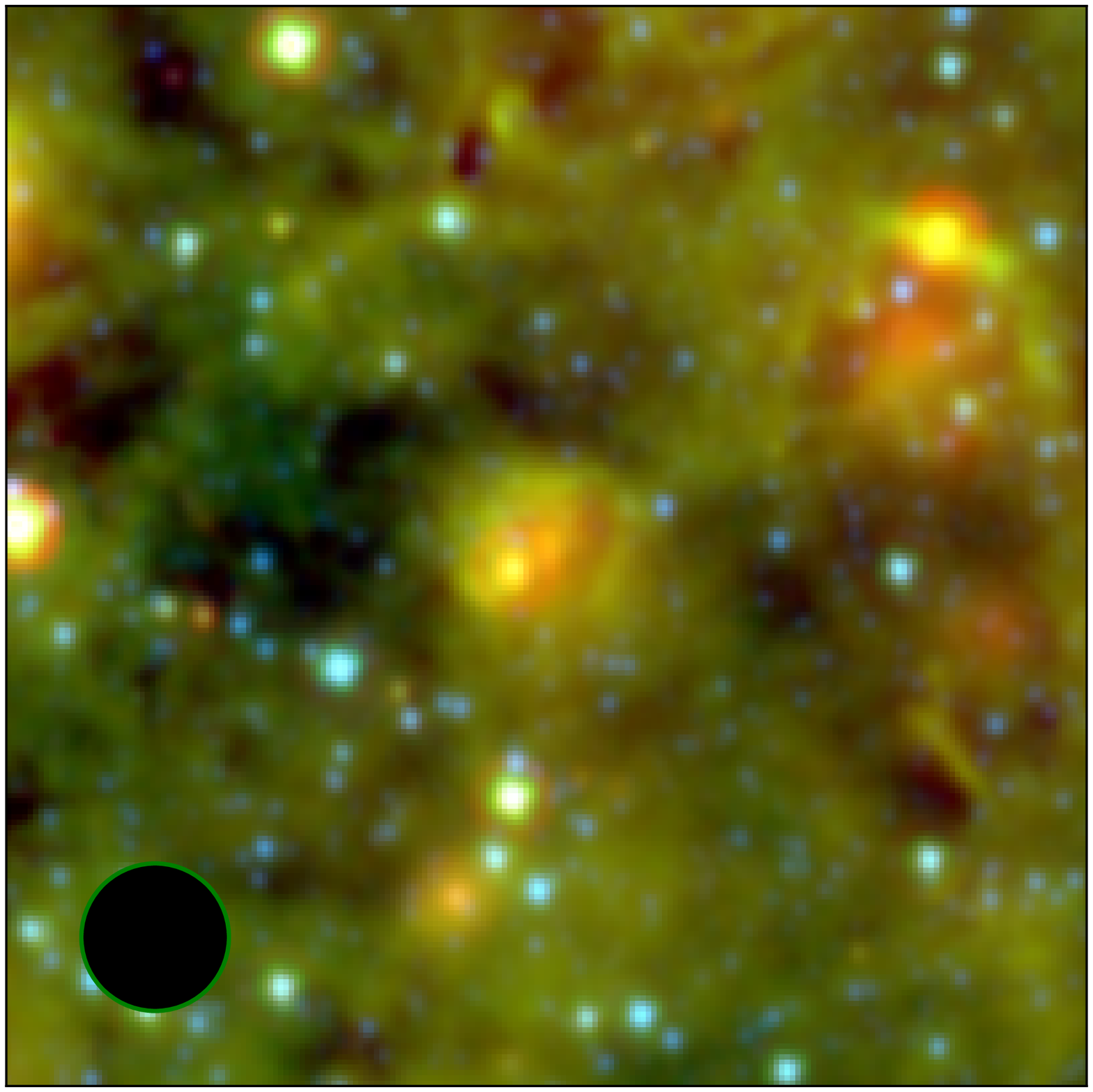}
\includegraphics[height=1.5in]{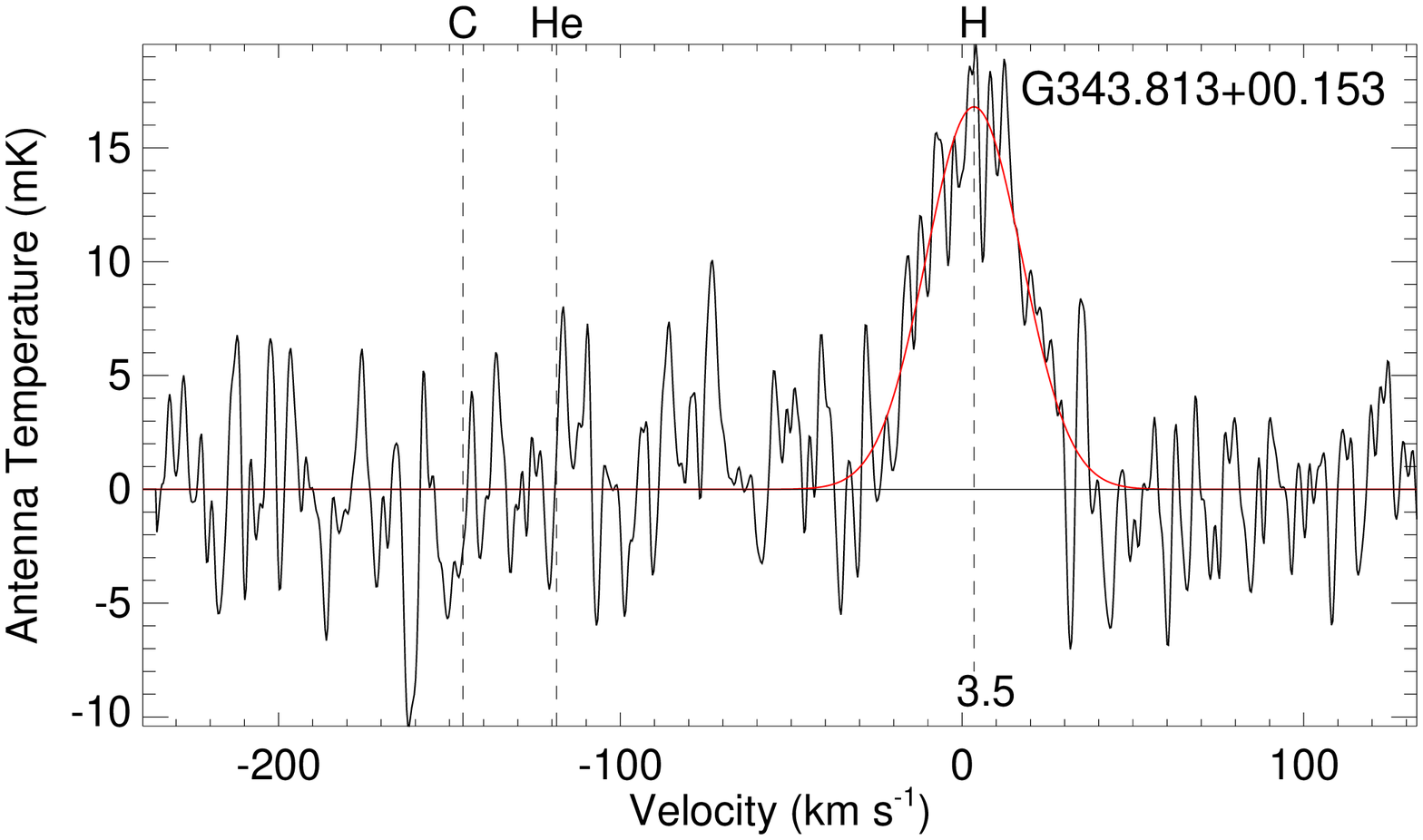}
\includegraphics[height=1.5in]{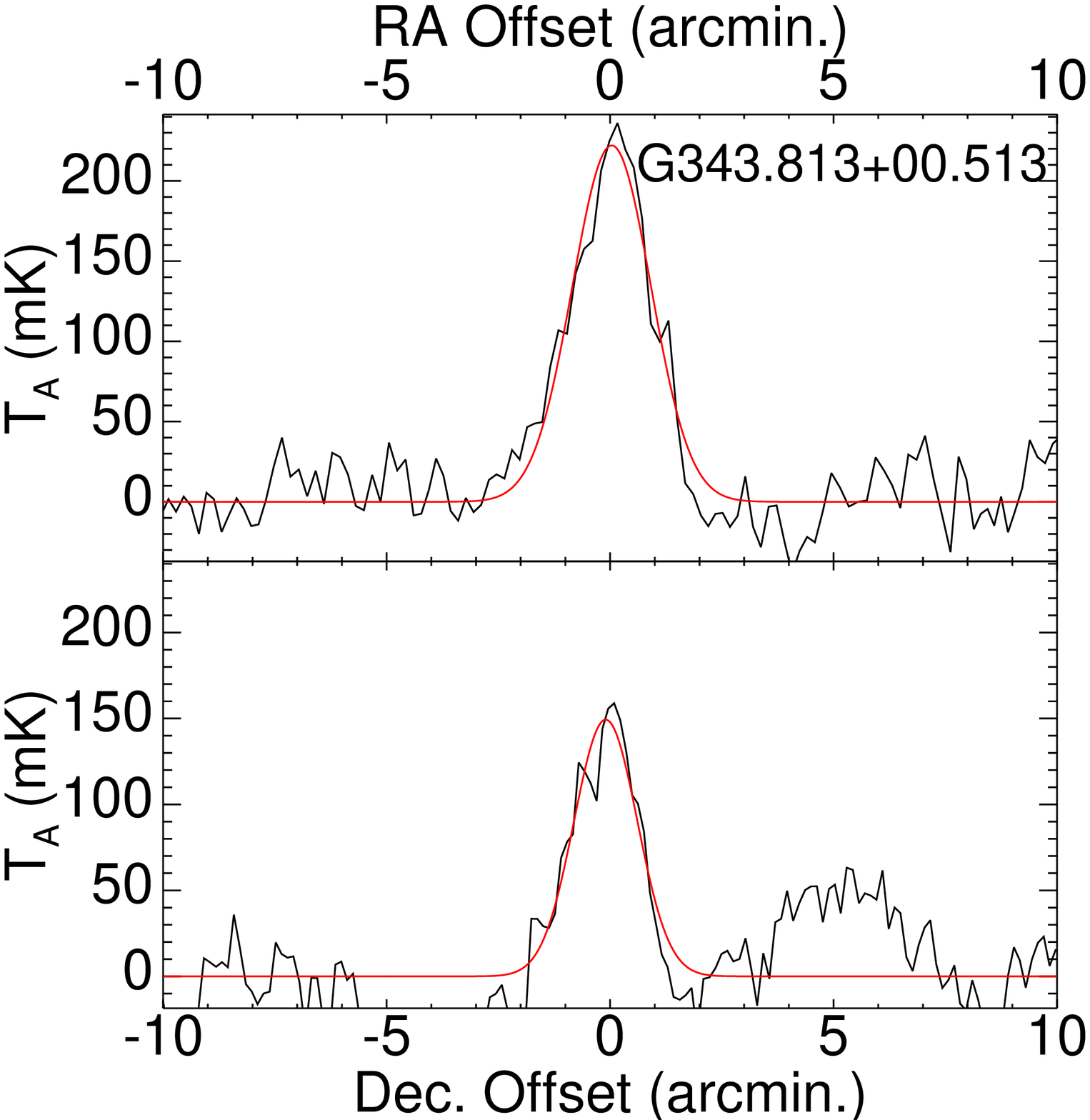}\\
\vskip 0.1in
\includegraphics[height=1.3in]{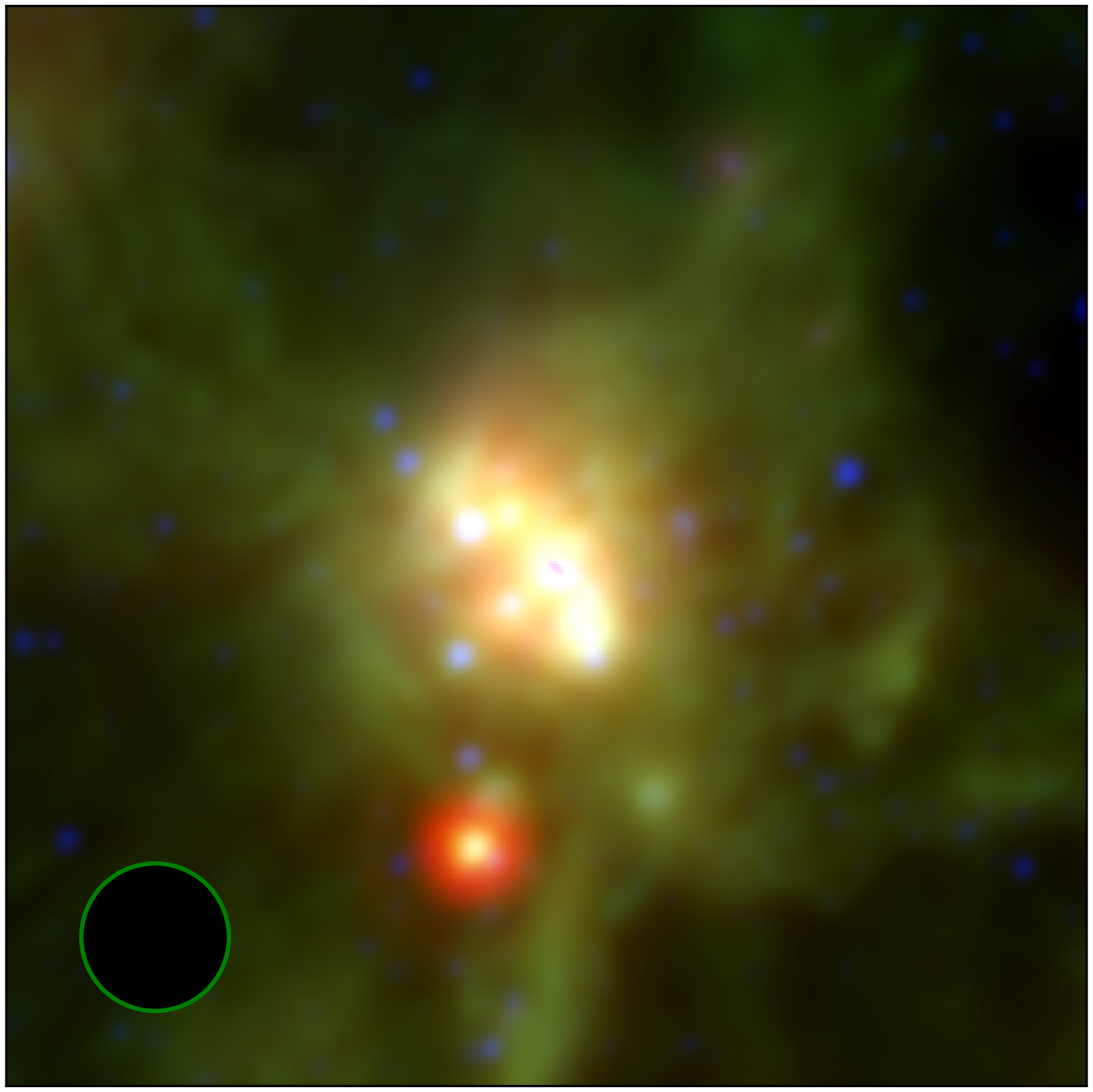}
\includegraphics[height=1.5in]{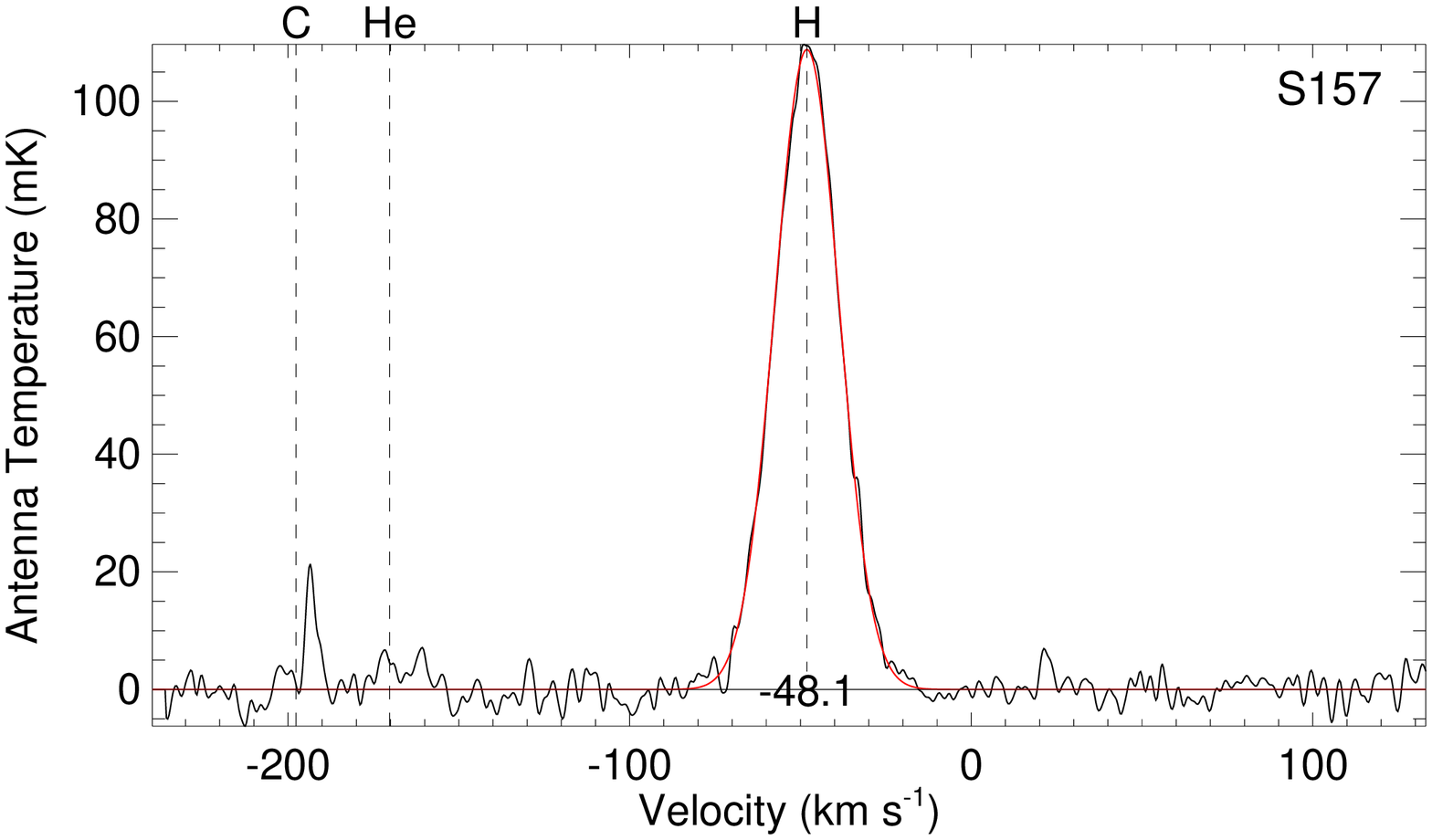}
\includegraphics[height=1.5in]{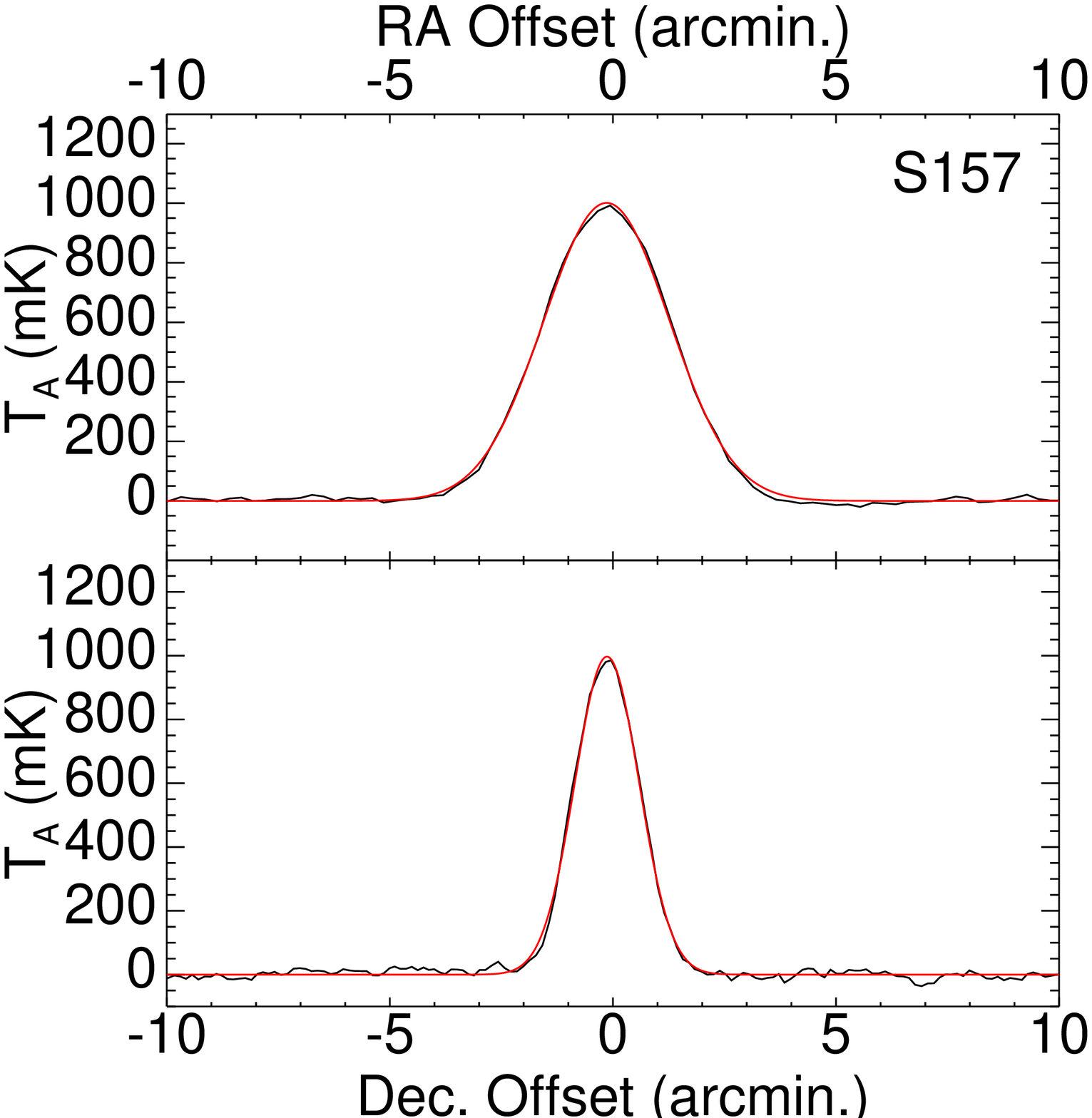}

\caption{Example \hii\ regions G033.006+01.151 (top row),
  G343.813+00.153 (middle) and S157 (bottom).  The left column shows
  WISE 22\,\micron\ emission in red, WISE 12\,\micron\ emission in
  green, and WISE 3.6\,\micron\ emission in blue.  Each image is
  $10\arcmin$ on a side oriented in Galactic coordinates and the
  $82\arcsec$ GBT beam is shown in the lower left.  \hii\ region
  candidates were identified based on their characteristic MIR
  morphology.  The middle column shows average X-band \hnaa\ spectra.
  The spectra have been smoothed to 1.86\,\kms\ resolution and the
  hydrogen Gaussian fits are shown in red.  The expected velocities of
  the helium and carbon RRLs are also indicated.  The right column
  shows continuum cross scans in RA (top panel) and Decl. (bottom
  panel).  Gaussian fits to the sources are shown in red.}
\label{fig:survey_examples}
\end{figure}

We have updated the GBT HRDS website with results from this
work\footnote{http://go.nrao.edu/hrds}.  This site contains images
such as those in Figure~\ref{fig:survey_examples} for all detected
sources, as well as the same for the sources from A11 and
\citet{bania12}.  We have also updated the WISE Catalog of Galactic
\hii\ Regions web site\footnote{http://astro.phys.wvu.edu/wise} with
results from these observations.

\begin{acknowledgments}
\nraoblurb\ This work supported by NASA ADAP grant NNX12AI59G and NSF
grant AST1516021.  We thank the staff at the Green Bank Telescope for
their hospitality and friendship during the observations and data
reduction.  We thank West Virginia University for its financial
support of GBT operations, which enabled some of the observations
for this project.

\it Facility: Green Bank Telescope
\end{acknowledgments}

\clearpage
\bibliographystyle{apj} 
\bibliography{../../ref.bib} 

\begin{thebibliography}{}
\expandafter\ifx\csname natexlab\endcsname\relax\def\natexlab#1{#1}\fi

\bibitem[{{Anderson} \& {Bania}(2009)}]{anderson09a}
{Anderson}, L.~D., \& {Bania}, T.~M. 2009, \apj, 690, 706

\bibitem[{{Anderson} {et~al.}(2014){Anderson}, {Bania}, {Balser}, {Cunningham},
  {Wenger}, {Johnstone}, \& {Armentrout}}]{anderson14}
{Anderson}, L.~D., {Bania}, T.~M., {Balser}, D.~S., {et~al.} 2014, \apjs, 212,
  1

\bibitem[{{Anderson} {et~al.}(2011){Anderson}, {Bania}, {Balser}, \&
  {Rood}}]{anderson11}
{Anderson}, L.~D., {Bania}, T.~M., {Balser}, D.~S., \& {Rood}, R.~T. 2011,
  \apjs, 194, 32

\bibitem[{{Anderson} {et~al.}(2012{\natexlab{a}}){Anderson}, {Bania}, {Balser},
  \& {Rood}}]{anderson12c}
---. 2012{\natexlab{a}}, \apj, 754, 62

\bibitem[{{Anderson} {et~al.}(2009){Anderson}, {Bania}, {Jackson}, {Clemens},
  {Heyer}, {Simon}, {Shah}, \& {Rathborne}}]{anderson09b}
{Anderson}, L.~D., {Bania}, T.~M., {Jackson}, J.~M., {et~al.} 2009, \apjs, 181,
  255

\bibitem[{{Anderson} {et~al.}(2015){Anderson}, {Hough}, {Wenger}, {Bania}, \&
  {Balser}}]{anderson15b}
{Anderson}, L.~D., {Hough}, L.~A., {Wenger}, T.~V., {Bania}, T.~M., \&
  {Balser}, D.~S. 2015, \apj, 810, 42

\bibitem[{{Anderson} {et~al.}(2012{\natexlab{b}}){Anderson}, {Zavagno},
  {Barlow}, {Garc{\'{\i}}a-Lario}, \& {Noriega-Crespo}}]{anderson12a}
{Anderson}, L.~D., {Zavagno}, A., {Barlow}, M.~J., {Garc{\'{\i}}a-Lario}, P.,
  \& {Noriega-Crespo}, A. 2012{\natexlab{b}}, \aap, 537, A1

\bibitem[{{Balser}(2006)}]{balser06}
{Balser}, D.~S. 2006, \aj, 132, 2326

\bibitem[{{Balser} {et~al.}(2011){Balser}, {Rood}, {Bania}, \&
  {Anderson}}]{balser11}
{Balser}, D.~S., {Rood}, R.~T., {Bania}, T.~M., \& {Anderson}, L.~D. 2011,
  \apj, 738, 27

\bibitem[{{Bania} {et~al.}(2012){Bania}, {Anderson}, \& {Balser}}]{bania12}
{Bania}, T.~M., {Anderson}, L.~D., \& {Balser}, D.~S. 2012, \apj, 759, 96

\bibitem[{{Bania} {et~al.}(2010){Bania}, {Anderson}, {Balser}, \&
  {Rood}}]{bania10}
{Bania}, T.~M., {Anderson}, L.~D., {Balser}, D.~S., \& {Rood}, R.~T. 2010,
  \apjl, 718, L106

\bibitem[{{Bock} {et~al.}(1999){Bock}, {Large}, \& {Sadler}}]{bock99}
{Bock}, D., {Large}, M.~I., \& {Sadler}, E.~M. 1999, \aj, 117, 1578

\bibitem[{{Brand} \& {Blitz}(1993)}]{brand93}
{Brand}, J., \& {Blitz}, L. 1993, \aap, 275, 67

\bibitem[{{Carey} {et~al.}(2009){Carey}, {Noriega-Crespo}, {Mizuno}, {Shenoy},
  {Paladini}, {Kraemer}, {Price}, {Flagey}, {Ryan}, {Ingalls}, {Kuchar},
  {Pinheiro Gon{\c c}alves}, {Indebetouw}, {Billot}, {Marleau}, {Padgett},
  {Rebull}, {Bressert}, {Ali}, {Molinari}, {Martin}, {Berriman}, {Boulanger},
  {Latter}, {Miville-Deschenes}, {Shipman}, \& {Testi}}]{carey09}
{Carey}, S.~J., {Noriega-Crespo}, A., {Mizuno}, D.~R., {et~al.} 2009, \pasp,
  121, 76

\bibitem[{{Choi} {et~al.}(2014){Choi}, {Hachisuka}, {Reid}, {Xu}, {Brunthaler},
  {Menten}, \& {Dame}}]{choi14}
{Choi}, Y.~K., {Hachisuka}, K., {Reid}, M.~J., {et~al.} 2014, \apj, 790, 99

\bibitem[{{Condon} {et~al.}(1998){Condon}, {Cotton}, {Greisen}, {Yin},
  {Perley}, {Taylor}, \& {Broderick}}]{condon98}
{Condon}, J.~J., {Cotton}, W.~D., {Greisen}, E.~W., {et~al.} 1998, \aj, 115,
  1693

\bibitem[{{Cram} {et~al.}(1996){Cram}, {Claussen}, {Beasley}, {Gray}, \&
  {Goss}}]{cram96}
{Cram}, L.~E., {Claussen}, M.~J., {Beasley}, A.~J., {Gray}, A.~D., \& {Goss},
  W.~M. 1996, \mnras, 280, 1110

\bibitem[{{Dame} \& {Thaddeus}(2011)}]{dame11}
{Dame}, T.~M., \& {Thaddeus}, P. 2011, \apjl, 734, L24+

\bibitem[{{Digel} {et~al.}(1994){Digel}, {de Geus}, \& {Thaddeus}}]{digel94}
{Digel}, S., {de Geus}, E., \& {Thaddeus}, P. 1994, \apj, 422, 92

\bibitem[{{Fich} {et~al.}(1990){Fich}, {Dahl}, \& {Treffers}}]{fich90}
{Fich}, M., {Dahl}, G.~P., \& {Treffers}, R.~R. 1990, \aj, 99, 622

\bibitem[{{Ghigo} {et~al.}(2001){Ghigo}, {Maddalena}, {Balser}, \&
  {Langston}}]{ghigo01}
{Ghigo}, F., {Maddalena}, R., {Balser}, D., \& {Langston}, G. 2001, GBT
  Commissioning Memo 10

\bibitem[{{Gordon}(1976)}]{gordon76}
{Gordon}, M.~A. 1976, in {Methods of Experimental Physics: Volume 12:
  Astrophysics, Part C: Radio Observations}, ed. M.~L. {Meeks} (Academic
  Press), 277--283

\bibitem[{{Gray} {et~al.}(1993){Gray}, {Whiteoak}, {Cram}, \& {Goss}}]{gray93}
{Gray}, A.~D., {Whiteoak}, J.~B.~Z., {Cram}, L.~E., \& {Goss}, W.~M. 1993,
  \mnras, 264, 678

\bibitem[{{Kobayashi} {et~al.}(2008){Kobayashi}, {Yasui}, {Tokunaga}, \&
  {Saito}}]{kobayashi08}
{Kobayashi}, N., {Yasui}, C., {Tokunaga}, A.~T., \& {Saito}, M. 2008, \apj,
  683, 178

\bibitem[{{Kuchar} \& {Clark}(1997)}]{kuchar97}
{Kuchar}, T.~A., \& {Clark}, F.~O. 1997, \apj, 488, 224

\bibitem[{{Liszt}(1992)}]{liszt92}
{Liszt}, H.~S. 1992, \apjs, 82, 495

\bibitem[{{Lockman}(1989)}]{lockman89}
{Lockman}, F.~J. 1989, \apjs, 71, 469

\bibitem[{{Lockman} {et~al.}(1996){Lockman}, {Pisano}, \& {Howard}}]{lockman96}
{Lockman}, F.~J., {Pisano}, D.~J., \& {Howard}, G.~J. 1996, \apj, 472, 173

\bibitem[{{McClure-Griffiths} {et~al.}(2005){McClure-Griffiths}, {Dickey},
  {Gaensler}, {Green}, {Haverkorn}, \& {Strasser}}]{mcclure05}
{McClure-Griffiths}, N.~M., {Dickey}, J.~M., {Gaensler}, B.~M., {et~al.} 2005,
  \apjs, 158, 178

\bibitem[{{Peng} {et~al.}(2000){Peng}, {Kraus}, {Krichbaum}, \&
  {Witzel}}]{peng00}
{Peng}, B., {Kraus}, A., {Krichbaum}, T.~P., \& {Witzel}, A. 2000, \aaps, 145,
  1

\bibitem[{{Quireza} {et~al.}(2006{\natexlab{a}}){Quireza}, {Rood}, {Balser}, \&
  {Bania}}]{quireza06a}
{Quireza}, C., {Rood}, R.~T., {Balser}, D.~S., \& {Bania}, T.~M.
  2006{\natexlab{a}}, \apjs, 165, 338

\bibitem[{{Quireza} {et~al.}(2006{\natexlab{b}}){Quireza}, {Rood}, {Bania},
  {Balser}, \& {Maciel}}]{quireza06b}
{Quireza}, C., {Rood}, R.~T., {Bania}, T.~M., {Balser}, D.~S., \& {Maciel},
  W.~J. 2006{\natexlab{b}}, \apj, 653, 1226

\bibitem[{{Reid} {et~al.}(2009){Reid}, {Menten}, {Zheng}, {Brunthaler}, \&
  {Xu}}]{reid09c}
{Reid}, M.~J., {Menten}, K.~M., {Zheng}, X.~W., {Brunthaler}, A., \& {Xu}, Y.
  2009, \apj, 705, 1548

\bibitem[{{Reid} {et~al.}(2014){Reid}, {Menten}, {Brunthaler}, {Zheng}, {Dame},
  {Xu}, {Wu}, {Zhang}, {Sanna}, {Sato}, {Hachisuka}, {Choi}, {Immer},
  {Moscadelli}, {Rygl}, \& {Bartkiewicz}}]{reid14}
{Reid}, M.~J., {Menten}, K.~M., {Brunthaler}, A., {et~al.} 2014, \apj, 783, 130

\bibitem[{{Russeil}(2003)}]{russeil03}
{Russeil}, D. 2003, \aap, 397, 133

\bibitem[{{Russeil} {et~al.}(2007){Russeil}, {Adami}, \&
  {Georgelin}}]{russeil07}
{Russeil}, D., {Adami}, C., \& {Georgelin}, Y.~M. 2007, \aap, 470, 161

\bibitem[{{Sharpless}(1959)}]{sharpless59}
{Sharpless}, S. 1959, \apjs, 4, 257

\bibitem[{{Stil} {et~al.}(2006){Stil}, {Taylor}, {Dickey}, {Kavars}, {Martin},
  {Rothwell}, {Boothroyd}, {Lockman}, \& {McClure-Griffiths}}]{stil06}
{Stil}, J.~M., {Taylor}, A.~R., {Dickey}, J.~M., {et~al.} 2006, \aj, 132, 1158

\bibitem[{{Sun} {et~al.}(2015){Sun}, {Xu}, {Yang}, {Li}, {Du}, {Zhang}, \&
  {Zhou}}]{sun15}
{Sun}, Y., {Xu}, Y., {Yang}, J., {et~al.} 2015, \apjl, 798, L27

\bibitem[{{Sunada} {et~al.}(2007){Sunada}, {Nakazato}, {Ikeda}, {Hongo},
  {Kitamura}, \& {Yang}}]{sunada07}
{Sunada}, K., {Nakazato}, T., {Ikeda}, N., {et~al.} 2007, \pasj, 59, 1185

\bibitem[{{Taylor} {et~al.}(2003){Taylor}, {Gibson}, {Peracaula}, {Martin},
  {Landecker}, {Brunt}, {Dewdney}, {Dougherty}, {Gray}, {Higgs}, {Kerton},
  {Knee}, {Kothes}, {Purton}, {Uyaniker}, {Wallace}, {Willis}, \&
  {Durand}}]{taylor03}
{Taylor}, A.~R., {Gibson}, S.~J., {Peracaula}, M., {et~al.} 2003, \aj, 125,
  3145

\bibitem[{{Taylor} \& {Cordes}(1993)}]{taylor93}
{Taylor}, J.~H., \& {Cordes}, J.~M. 1993, \apj, 411, 674

\bibitem[{{Wenger} {et~al.}(2013){Wenger}, {Bania}, {Balser}, \&
  {Anderson}}]{wenger13}
{Wenger}, T.~V., {Bania}, T.~M., {Balser}, D.~S., \& {Anderson}, L.~D. 2013,
  \apj, 764, 34

\bibitem[{{Yasui} {et~al.}(2008){Yasui}, {Kobayashi}, {Tokunaga}, {Terada}, \&
  {Saito}}]{yasui08}
{Yasui}, C., {Kobayashi}, N., {Tokunaga}, A.~T., {Terada}, H., \& {Saito}, M.
  2008, \apj, 675, 443

\end{thebibliography}

\clearpage

\begin{deluxetable}{lccrcccrccc}
\tabletypesize{\scriptsize}
\tablecaption{Hydrogen Recombination Line Parameters}
\tablewidth{0pt}
\tablehead{
\colhead{Source} &
\colhead{\gl} &
\colhead{\gb} &
\colhead{$T_L$} &
\colhead{$\sigma T_L$} &
\colhead{$\Delta V$} &
\colhead{$\sigma \Delta V$} &
\colhead{$V_{LSR}$} &
\colhead{$\sigma V_{LSR}$} &
\colhead{r.m.s.} &
\colhead{Note\tablenotemark{a}}
\\
\colhead{} &
\colhead{$\arcdeg$} &
\colhead{$\arcdeg$} &
\colhead{mK} &
\colhead{mK} &
\colhead{\kms} &
\colhead{\kms} &
\colhead{\kms} &
\colhead{\kms} &
\colhead{mK} &
\colhead{}
}
\startdata

G000.335$-$00.015a &    0.335 & $-$0.015 & $\phn 164.9$  &  $1.7$  &  $33.7$  &  $0.5$  &  \phn 14.0  &  $0.3$ & \phn 3.1 & * \\
G000.335$-$00.015b &    0.335 & $-$0.015 & $\phn \phn 47.5$  &  $1.8$  &  $32.2$  &  $1.8$  &  \phn 46.5  &  $1.1$ & \phn 3.1 & \\
G000.492+00.192 &    0.492 & +0.192 & $\phn 106.0$  &  $0.5$  &  $22.1$  &  $0.1$  &  \phn \phn $-7.5$  &  $0.1$ & \phn 2.6 & \\
G000.583$-$00.870 &    0.583 & $-$0.870 & $\phn 281.4$  &  $2.2$  &  $17.4$  &  $0.2$  &  \phn 14.2  &  $0.1$ & \phn 2.4 & \\
G000.602$-$00.051 &    0.602 & $-$0.051 & $\phn 568.2$  &  $2.5$  &  $41.1$  &  $0.2$  &  \phn 53.1  &  $0.1$ & \phn 2.3 & \\
G000.630$-$00.027a &    0.630 & $-$0.027 & $\phn 188.2$  &  $0.3$  &  $52.8$  &  $0.5$  &  \phn 90.6  &  $0.1$ & \phn 3.3 & \\
G000.630$-$00.027b &    0.630 & $-$0.027 & $\phn \phn 87.2$  &  $0.7$  &  $37.0$  &  $0.4$  &  \phn 51.3  &  $0.2$ & \phn 3.3 & \\
G000.630$-$00.027c &    0.630 & $-$0.027 & $\phn \phn 27.9$  &  $0.5$  &  $28.8$  &  $0.6$  &  136.2  &  $0.3$ & \phn 3.3 & \\
G000.639$-$00.114a &    0.639 & $-$0.114 & $\phn 192.6$  &  $1.1$  &  $27.4$  &  $0.4$  &  \phn 70.4  &  $0.1$ & \phn 3.3 & \\
G000.639$-$00.114b &    0.639 & $-$0.114 & $\phn \phn 65.3$  &  $1.1$  &  $27.6$  &  $0.6$  &  \phn 24.9  &  $0.2$ & \phn 3.3 & \\
G000.639$-$00.114c &    0.639 & $-$0.114 & $\phn \phn 33.4$  &  $1.7$  &  $17.4$  &  $1.2$  &  \phn 96.7  &  $0.6$ & \phn 3.3 & \\
G000.670$-$00.034 &    0.670 & $-$0.034 & $2047.0$  &  $5.9$  &  $40.2$  &  $0.1$  &  \phn 64.8  &  $0.1$ & 11.7 & \\
G000.682$-$00.035 &    0.682 & $-$0.035 & $1804.1$  &  $5.0$  &  $39.1$  &  $0.1$  &  \phn 66.4  &  $0.1$ & \phn 5.7 & \\
G000.692$-$00.046 &    0.692 & $-$0.046 & $\phn 504.5$  &  $1.2$  &  $40.9$  &  $0.1$  &  \phn 60.0  &  $0.1$ & \phn 3.9 & \\
G001.005$-$00.236 &    1.005 & $-$0.236 & $\phn \phn 32.2$  &  $0.4$  &  $22.4$  &  $0.3$  &  \phn \phn 2.5  &  $0.1$ & \phn 2.4 & \\

\enddata
\label{tab:line}
\tablenotetext{a}{Discrete HII region velocity for sources with multiple velocity component spectra marked with an asterisk (*).}
\tablecomments{Table~\ref{tab:line} is available in its entirety in the electronic edition of the {\it Astrophysical Journal Supplement Series}. 
A portion is shown here for guidance regarding its form and content.}
\end{deluxetable}
\clearpage


\begin{deluxetable}{lccccccccccccc}
\tabletypesize{\scriptsize}
\tablecaption{Radio Continuum Parameters}
\tablewidth{0pt}
\tablehead{
\colhead{Source} &
\colhead{\gl} &
\colhead{\gb} &
\colhead{$T_\alpha$} &
\colhead{$\sigma T_\alpha$} &
\colhead{$T_\delta$} &
\colhead{$\sigma T_\delta$} &
\colhead{$\Theta_\alpha$} &
\colhead{$\sigma\Theta_\alpha$} &
\colhead{$\Theta_\delta$} &
\colhead{$\sigma\Theta_\delta$} &
\colhead{$S$} &
\colhead{$\sigma S$} &
\colhead{Note\tablenotemark{a}}\\
\colhead{} &
\colhead{$\arcdeg$} &
\colhead{$\arcdeg$} &
\colhead{mK} &
\colhead{mK} &
\colhead{mK} &
\colhead{mK} &
\colhead{$\arcsec$} &
\colhead{$\arcsec$} &
\colhead{$\arcsec$} &
\colhead{$\arcsec$} &
\colhead{mJy} &
\colhead{mJy} &
\colhead{}
}
\startdata

G000.335$-$00.015 & $    0.335$  &  $-0.015$  &  \phn 2101  &  \phn 17  &  \phn 2109  &  \phn 29  &  129  &  \phn \phn 2  &  111  &  \phn 3  &  \phn 2254  &  \phn 177  &  P \\
G000.492+00.192 & $    0.492$  &  $+0.192$  &  \phn \phn 673  &  \phn 13  &  \phn \phn 628  &  \phn \phn 7  &  134  &  \phn \phn 5  &  105  &  \phn 2  &  \phn \phn 680  &  \phn \phn 67  &   \\
G000.583$-$00.870 & $    0.583$  &  $-0.870$  &  \phn \phn 765  &  \phn \phn 6  &  \phn 1380  &  \phn 11  &  144  &  \phn \phn 4  &  108  &  \phn 2  &  \phn 1241  &  \phn 101  &  C \\
G000.602$-$00.051 & $    0.602$  &  $-0.051$  &  \phn 7709  &  \phn 86  &  \phn 7947  &  \phn 55  &  173  &  \phn \phn 3  &  135  &  \phn 2  &  13611  &  \phn 940  &  P \\
G000.630$-$00.027 & $    0.630$  &  $-0.027$  &  \nodata  &  \nodata  &  \phn 7426  &  \phn 16  &  \nodata  &  \nodata  &  119  &  \phn 1  &  10779  &  \phn 106  &   \\
G000.639$-$00.114 & $    0.639$  &  $-0.114$  &  \nodata  &  \nodata  &  \phn 2116  &  \phn 49  &  \nodata  &  \nodata  &  116  &  \phn 2  &  \phn 3004  &  \phn 118  &   \\
G000.670$-$00.034 & $    0.670$  &  $-0.034$  &  34987  &  218  &  35718  &  225  &  104  &  \phn \phn 1  &  116  &  \phn 2  &  31887  &  1495  &  P \\
G000.682$-$00.035 & $    0.682$  &  $-0.035$  &  30708  &  448  &  30434  &  164  &  113  &  \phn \phn 3  &  117  &  \phn 1  &  30075  &  1882  &   \\
G000.692$-$00.046 & $    0.692$  &  $-0.046$  &  \nodata  &  \nodata  &  \phn 6560  &  \phn 62  &  \nodata  &  \nodata  &  115  &  \phn 3  &  \phn 9173  &  \phn 252  &   \\
G001.005$-$00.236 & $    1.005$  &  $-0.236$  &  \phn \phn 481  &  \phn \phn 8  &  \phn \phn 458  &  \phn 12  &  178  &  \phn \phn 6  &  147  &  \phn 8  &  \phn \phn 911  &  \phn 162  &   \\

\enddata
\tablenotetext{a}{Comments concerning continuum emission morphology and data 
quality.\\
C -- Complex:  Source has complex continuum structure; this single Gaussian 
component model only crudely represents the true source characteristics 
(see text).\\ 
P -- Peaked:  Continuum peak lies within $10\arcsec$ of the nominal target position. 
These are the highest quality continuum data.}
\label{tab:continuum}
\tablecomments{Table~\ref{tab:continuum} is available in its entirety in the electronic edition of the {\it Astrophysical Journal Supplement Series}. 
A portion is shown here for guidance regarding its form and content.}
\end{deluxetable}
\clearpage


\begin{deluxetable}{lccccc}
\tablecaption{Spectrophotometric and Kinematic Distance Comparison}
\tablewidth{0pt}
\tablehead{
\colhead{Name} &
\colhead{$d_{\rm Spec}$} &
\colhead{$\sigma d_{\rm Spec}$} &
\colhead{$d_{\rm Kin}$} &
\colhead{$\sigma d_{\rm Kin}$} &
\colhead{Reference$^{\rm a}$}
}
\startdata
S47	&	2.8	&	0.6	&	\phn3.1	&	0.4	&	1	\\
S57	&	1.7	&	0.5	&	\phn1.5	&	0.5	&	1	\\
S61	&	2.3	&	0.7	&	\phn2.3	&	0.4	&	1	\\
S69	&	3.8	&	1.1	&	\phn3.6	&	0.4	&	1	\\
S92	&	3.9	&	0.5	&	\phn3.7	&	0.1	&	1	\\
S107	&	2.6	&	0.8	&	\phn1.9	&	0.4	&	1	\\
S120	&	6.3	&	1.1	&	10.0	&	1.3	&	1	\\
S121	&	6.3	&	1.1	&	\phn8.5	&	1.2	&	1	\\
S124	&	3.6	&	0.8	&	\phn5.6	&	1.4	&	1	\\
S127	&	9.7	&	2.9	&	12.9	&	1.7	&	1	\\
S128	&	7.8	&	2.3	&	\phn9.2	&	1.5	&	1	\\
S135	&	1.2	&	0.3	&	\phn3.1	&	1.3	&	1	\\
S141	&	8.3	&	0.6	&	\phn7.0	&	1.2	&	2	\\
S148	&	5.3	&	1.1	&	\phn5.8	&	1.3	&	1	\\
S157	&	3.2	&	0.3	&	\phn5.1	&	1.4	&	1	\\
S164	&	2.4	&	0.4	&	\phn4.8	&	1.3	&	1	\\
S168	&	2.5	&	0.5	&	\phn4.4	&	1.3	&	1	\\
S170	&	2.6	&	0.6	&	\phn5.8	&	1.1	&	2	\\
S175	&	2.3	&	0.7	&	\phn5.4	&	1.1	&	2	\\
S193	&	2.4	&	0.3	&	\phn5.2	&	1.3	&	2	\\
S207	&	7.5	&	1.6	&	\phn4.5	&	1.5	&	1	\\
S208	&	7.5	&	1.6	&	\phn3.9	&	1.4	&	1	\\
S219	&	4.0	&	0.7	&	\phn4.9	&	2.1	&	1	\\
S294	&	3.2	&	1.6	&	\phn3.6	&	1.2	&	1	

\enddata
\tablenotetext{a}{1 = \citet{russeil03}; 2 = \citet{russeil07}}
\label{tab:sharpless}
\end{deluxetable}
\clearpage


\begin{deluxetable}{lcc}
\tablecaption{HII Region Kinematic Distance Summary$^{\rm a}$}
\tablewidth{0pt}
\tablehead{
\colhead{Category} &
\colhead{Previously Known} &
\colhead{This Work}
}
\startdata
{\bf Inner Galaxy ($\rgal < 8.5\,\kpc$)}				&	{\bf 661}	&\phn	 {\bf 26}	\\
\,\,\,\,\,\,\,\,\,\,\,\,\,\,Sharpless (Near Distance)	 			&	\phn17 	&\phn\phn	  4	\\
\,\,\,\,\,\,\,\,\,\,\,\,\,\,Tangent Point			&	166  	&\phn\phn	  6	\\
\,\,\,\,\,\,\,\,\,\,\,\,\,\,KDAR	 			&	478 	&\phn	 16	\\
{\bf Outer Galaxy ($\rgal > 8.5\,\kpc$)}				&	{\bf 160}	&	{\bf 105}	\\
\,\,\,\,\,\,\,\,\,\,\,\,\,\,1st Quadrant			&	\phn54 	&\phn	 34	\\
\,\,\,\,\,\,\,\,\,\,\,\,\,\,2nd Quadrant			&	\phn72	&\phn	 66	\\
\,\,\,\,\,\,\,\,\,\,\,\,\,\,3rd Quadrant ($\ell < 225\degree$)	&	\phn27	& \phn\phn	  1	\\
\,\,\,\,\,\,\,\,\,\,\,\,\,\,4th Quadrant ($\ell > 340\degree$)	&   \phn\phn 7 	&	\phn\phn  4	\\
{\bf Total}					 	  		&       {\bf 821}  	&	{\bf 131}	\\
\enddata
\tablenotetext{a}{Table only includes \hii\ regions with known distances.  The ``Previously Known'' values are from A14.}
\label{tab:distance_summary}
\end{deluxetable}
\clearpage

\begin{deluxetable}{lrcccccccrrr}
\setlength{\tabcolsep}{6pt}
\tabletypesize{\footnotesize}
\tablecaption{Kinematic Distances}
\tablewidth{0pt}
\tablehead{
\colhead{Name} &
\colhead{$V_{\rm LSR}$} &
\colhead{$D_N$} &
\colhead{$D_F$} &
\colhead{$D_T$} &
\colhead{\rgal} &
\colhead{$V_{\rm T}$} &
\colhead{KDAR$^{\rm a}$} &
\colhead{$D_\sun$} &
\colhead{$\sigma D\sun$} &
\colhead{$z$}
\\
\colhead{} &
\colhead{\kms} &
\colhead{kpc} &
\colhead{kpc} &
\colhead{kpc} &
\colhead{kpc} &
\colhead{\kms} &
\colhead{} &
\colhead{kpc} &
\colhead{kpc} &
\colhead{pc}
}
\startdata
G010.621$-$00.318 & \phn $-10.0$ & \nodata & 19.8 & 8.4 & 11.6 & 168.7 & O & 19.8 & 3.1  & $-110$\\
G010.630$-$00.338 & \phn \phn $-9.1$ & \nodata & 19.5 & 8.4 & 11.2 & 168.7 & O & 19.5 & 3.0  & $-114$\\
S47 & $\phn 69.2$ & 5.1 & 11.3 & 8.2 & \phn 3.8 & 154.4 & N & \phn 5.1 & 0.4 & \phn 16\\
G019.716$-$00.261 & $\phn 40.1$ & 3.3 & 12.7 & 8.0 & \phn 5.5 & 139.9 & F & 12.7 & 0.4 & \phn $-57$\\
G021.603$-$00.169a & \phn \phn $-4.7$ & \nodata & 16.5 & 7.9 & \phn 9.2 & 133.9 & O & 16.5 & 1.1 & \phn $-48$\\
S57 & $\phn 31.0$ & 2.5 & 13.1 & 7.8 & \phn 6.3 & 129.8 & N & \phn 2.5 & 0.5 & \phn 29\\
G023.295$-$00.280b & $\phn 61.6$ & 4.2 & 11.4 & 7.8 & \phn 4.9 & 128.4 & F & 11.4 & 0.6 & \phn $-55$\\
G025.520+00.215a & $\phn 37.4$ & 2.8 & 12.6 & 7.7 & \phn 6.1 & 121.4 & F & 12.6 & 0.7 & \phn 47\\
G025.651$-$00.031 & $\phn 87.5$ & 5.3 & 10.0 & 7.7 & \phn 4.4 & 121.0 & F & 10.0 & 0.4 & \phn \phn $-5$\\
S61 & $\phn 37.3$ & 2.7 & 12.5 & 7.6 & \phn 6.2 & 118.4 & N & \phn 2.7 & 0.4 & \phn 83\\
\enddata
\tablenotetext{a}{Kinematic distance ambiguity resolution: ``N''=near distance; ``F''=far distance; ``O''=outer Galaxy source}
\tablecomments{Table~\ref{tab:distances} is available in its entirety in the electronic edition of the {\it Astrophysical Journal Supplement Series}. 
A portion is shown here for guidance regarding its form and content.}
\label{tab:distances}
\end{deluxetable}

\clearpage

\begin{deluxetable}{lrrrrrrrrrrrr}
\tabletypesize{\small}
\tablecaption{OSC HII Regions?}
\tablewidth{0pt}
\tablehead{
\colhead{} &
\colhead{} &
\colhead{} &
\colhead{} &
\colhead{} &
\multicolumn{3}{c}{Brand} &
\colhead{} &
\multicolumn{3}{c}{Reid}
\\ \cline{6-8} \cline{10-12}
\colhead{Name} &
\colhead{\gl} &
\colhead{\gb} &
\colhead{$V_{\rm LSR}$} &
\colhead{} &
\colhead{\rgal} &
\colhead{$d_\sun$} &
\colhead{$z$} &
\colhead{} &
\colhead{\rgal} &
\colhead{$d_\sun$} &
\colhead{$z$}
\\
\colhead{} &
\colhead{$\arcdeg$} &
\colhead{$\arcdeg$} &
\colhead{\kms} &
\colhead{} &
\colhead{kpc} &
\colhead{kpc} &
\colhead{pc} &
\colhead{} &
\colhead{kpc} &
\colhead{kpc} &
\colhead{pc}
}
\startdata
G033.008+01.151 & 33.008 & 1.151 & $-$57.6 & & 17.0 & 23.5 & 472 & & 14.8 & 21.1 &          424\\
G039.183$-$01.422 & 39.183 & $-$1.422 & $-$54.9 & & 14.5 & 20.1 & $-499$ & & 13.0 & 18.4 &         $-$456\\
G041.755+01.451 & 41.755 & 1.451 & $-$54.8 & & 14.0 & 19.2 & 485 & & 12.6 & 17.6 &          445\\
G041.804+01.503 & 41.804 & 1.503 & $-$52.6 & & 13.7 & 18.8 & 493 & & 12.4 & 17.3 &          453\\
G054.094+01.749 & 54.094 & 1.749 & $-$85.3 & & 17.0 & 20.5 & 626 & & 14.8 & 18.0 &          550\\
S83 & 55.114 & 2.422 & $-$76.1 & & 15.3 & 18.4 & 778 & & 13.5 & 16.5 &          695\\
\hdashline
G149.746$-$00.199 & 149.746 & $-$0.199 & $-$71.4 & & 25.3 & 17.5 & \phn $-60$ & & 20.2 & 12.5 & \phn         $-$43\\
G151.561$-$00.425 & 151.561 & $-$0.425 & $-$56.4 & & 19.3 & 11.4 & \phn $-84$ & & 16.4 & 8.5 & \phn         $-$63\\
G151.626$-$00.456 & 151.626 & $-$0.456 & $-$58.2 & & 20.1 & 12.2 & \phn $-97$ & & 16.9 & 9.1 & \phn         $-$72\\

\enddata
\label{tab:osc}
\end{deluxetable}

\clearpage

\clearpage

\clearpage


\clearpage

\clearpage


\clearpage

\end{document}